\documentclass[11pt]{llncs}

\usepackage{microtype}
\usepackage{amsmath}
\usepackage{graphicx}
\usepackage[boxed]{algorithm2e}
\DeclareGraphicsExtensions{.pdf,.png}
\usepackage{fullpage}
\usepackage{color}

\newcommand{\no}[1]{}

\newcommand{\removed}[1]{}

\newcommand{\F}{\mathcal{F}}

\newcommand{\Rel}{\mathcal{R}}
\newcommand{\G}{\mathcal{G}}
\newcommand{\X}{\mathcal{X}}
\newcommand{\Tree}{\mathcal{T}}

\newcommand{\Lab}{\mathcal{L}}

\newcommand{\node}{\mathit{node}}
\newcommand{\preorder}{\mathit{preorder}}
\newcommand{\leafrank}{\mathit{leafrank}}
\newcommand{\leafselect}{\mathit{leafselect}}
\newcommand{\intrank}{\mathit{intrank}}
\newcommand{\intselect}{\mathit{intselect}}
\newcommand{\numleaves}{\mathit{numleaves}}
\newcommand{\parent}{\mathit{parent}}
\newcommand{\child}{\mathit{child}}
\newcommand{\nextsibling}{\mathit{nextsibling}}

\newcommand{\degree}{\mathit{degree}}
\newcommand{\depth}{\mathit{depth}}
\newcommand{\levelanc}{{\mathit{level}\textrm{-}\mathit{ancestor}}}

\newcommand{\rank}{\mathit{rank}}
\newcommand{\select}{\mathit{select}}
\newcommand{\access}{\mathit{access}}

\title{Grammar-Compressed Indexes \\ with Logarithmic Search Time
\thanks{A preliminary version of this article appeared in {\em Proc. SPIRE'12}
\cite{CN12}. This work was partially funded by the Millennium Institute for
Foundational Research on Data (IMFD) and Fondecyt Grant 1-200038, Chile.}}

\author{Francisco Claude \inst{1}
\and
Gonzalo Navarro \inst{2,3}
\and
Alejandro Pacheco \inst{3}
}

\institute{
LinkedIn, USA
\and
Millennium Institute for Foundational Research on Data (IMFD), Chile
\and
Department of Computer Science, University of Chile}

\begin{document}

\maketitle

\begin{abstract} 
Let a text $T[1..n]$ be the only string generated by a context-free grammar 
with $g$ (terminal and nonterminal) symbols, and of size $G$ (measured as the 
sum of the lengths of the right-hand sides of the rules). Such a grammar, called
a grammar-compressed representation of $T$, can be encoded using
essentially $G\lg g$ bits. We introduce the first grammar-compressed 
{\em index} that uses $O(G\lg n)$ bits and can find the $occ$ occurrences of 
patterns $P[1..m]$ in time $O((m^2+occ)\lg G)$. We implement the index and
demonstrate its practicality in comparison with the state of the art, on highly
repetitive text collections.
\end{abstract}

\sloppy

\section{Introduction and Related Work}

Grammar-based compression is an active area of research since at
least the seventies \cite{CRA76,Sto77,ZL78,SS82}. A given sequence $T[1..n]$ 
over alphabet $[1..\sigma]$ is
replaced by a hopefully small (context-free) grammar $\G$ that generates just
the string $T$. Let $g$ be the number of grammar symbols, counting terminals
and nonterminals. Let $G=|\G|$ be the {\em size} of the grammar, measured as
the sum of the lengths of the right-hand sides of the rules. Then a basic 
grammar-compressed representation of $T$ requires essentially $G\lg g$ bits, 
instead of the $n\lg\sigma$ bits required by a plain representation. 
It always holds $G \ge \lg n$, and indeed $G$ can be as small as $O(\lg n)$ 
in extreme cases; consider $T=a^n$.

Grammar-based methods can achieve universal compression \cite{KY00}. Unlike
statistical methods, which exploit frequencies to achieve compression, 
grammar-based methods exploit repetitions in the text, and thus they are 
especially suitable for compressing highly repetitive sequence collections.  
These collections, containing long identical substrings that are possibly far 
away from each
other, arise when managing software repositories, versioned documents,
transaction logs, periodic publications, and computational biology sequence
databases, among others. Statistical compression is helpless to exploit this sort of
long-range repetitiveness \cite{KN12,Nav12}.

Finding the smallest grammar $\G^*$ that represents a given text $T$ is 
NP-complete \cite{Sto77,SS82,Ryt03,CLLPPSS05}. Moreover, the size $G^*$ of the 
smallest grammar is never smaller than the number $z$ of phrases in a Lempel-Ziv
parse \cite{LZ76} of $T$. A simple method to achieve an $O(\lg n)$-approximation
to the smallest grammar size is to parse $T$ using Lempel-Ziv and then convert it 
into a grammar \cite{Ryt03}. More sophisticated approximations 
\cite{Ryt03,CLLPPSS05,Jez15,Jez16} achieve ratio $O(\lg(n/G^*))$ (indeed,
they obtain size $G = O(z \lg(n/z))$). Recently, it has been shown that this
is not far from the lower bound, as there are sequence families where 
$G^* = \Omega(z \lg n / \lg\lg n)$ \cite{HLR16}.

The known approximation ratios of popular grammar compressors such as LZ78 
\cite{ZL78}, Re-Pair \cite{LM00} and Sequitur \cite{NMWM04}, instead, are much 
larger than the optimal \cite{CLLPPSS05,HLR16}. Still, some of those methods 
(in particular Re-Pair) perform very well in practice, both in classical and 
repetitive settings.\footnote{See the statistics in
{\tt http://pizzachili.dcc.uchile.cl/repcorpus.html}.}

On the other hand, unlike Lempel-Ziv, grammar compression allows one to decompress
arbitrary substrings of $T$ in logarithmic time \cite{GKPS05,BLRSRW15,BPT15}. 
The most recent results extract any $T[p..p+\ell-1]$ in time $O(\ell+\lg n)$
\cite{BLRSRW15} and even $O(\ell/\lg_\sigma n + \lg n)$ \cite{BPT15}, which
is close to optimal \cite{VY13}. 
Unfortunately, those representations require $O(G\lg n)$ bits, possibly 
proportional but in practice many times the size of the output of a grammar 
compressor.

More ambitious than just extracting substrings from $T$ is to ask for {\em
indexed searches}, that is, finding the $occ$ occurrences in $T$ of a given 
pattern $P[1..m]$. {\em Self-indexes} are compressed text representations that 
support both operations, {\em extract} $T[p..p+\ell-1]$ and {\em locate} the
occurrences of a pattern
$P[1..m]$, in time sublinear (and usually polylogarithmic) in $n$.
They have appeared in the last decade \cite{NM07},
and have focused mostly on statistical compression. As a result, they work
well on classical texts, but not on repetitive collections \cite{MNSV09}. Some
of those self-indexes have been adapted to such repetitive collections
\cite{MNSV09,NPLHLMP13,NPCHIMP13,DJSS14,BGGMS14,BCGPR15,GNP18}, but they do not 
reach the compression ratio of the best grammar-based methods. 


Searching for patterns on grammar-compressed text has been faced mostly in
sequential form \cite{AB92}, that is, scanning the whole grammar. The best
result \cite{KMSTSA03} achieves time $O(G+m^2+occ)$. This may be $o(n)$, but
is still linear in the size of the compressed text. There exist a few 
self-indexes based on LZ78-like compression \cite{FM05,RO08,ANS10}, but 
LZ78 is among the weakest grammar-based compressors. In particular, LZ78 has 
been shown not to be competitive on highly repetitive 
collections~\cite{MNSV09}.

The only self-index supporting general grammar compressors \cite{CN11}
operates on ``straight-line programs'' (SLPs), where the right hands of the
rules are of length 1 or 2. Given such a grammar they achieve, among other
tradeoffs, $3g\lg g + g\lg n$ bits of space and $O(m(m+h)\lg^2 g)$ search
time, where $h \le g$ is the height of the parse tree of the grammar. A general
grammar of $g$ symbols and size $G$ can be converted into an SLP by adding at 
most $G-2g$ symbols and/or rules. 

More recently, a self-index based on Lempel-Ziv compression was developed
\cite{KN12}. It uses
$z\lg z + 2z\lg n + O(z\lg\sigma)$ bits of space and searches in time $O(m^2
\bar{h} + (m+occ)\lg z)$, where $\bar{h} \le z$ is the nesting of the parsing.
Extraction requires $O(\ell\, \bar{h})$ time. Experiments on repetitive 
collections \cite{CFMPN10,CFMN16} showed that the grammar-based compressor 
\cite{CN11} can outperform the (by then) best classical self-index adapted to 
repetitive collections \cite{MNSV09} but, at least that particular 
implementation, was not
competitive with the Lempel-Ziv-based self-index \cite{KN12}.

The search times in both self-indexes depend on $h$ or $\bar{h}$. This is 
undesirable as both are only bounded by $g$ or $z$, respectively. As
mentioned, this kind of dependence has been removed for extracting text 
substrings \cite{BLRSRW15}, at the cost of using $O(G\lg n)$ further bits.

There have also been combinations of grammar-based and Lempel-Ziv-based methods 
\cite{GGKNP12,GGKNP14,BEGV18,CE18}, yet (1) none of those is implemented, (2)
the constant factors multiplying their space complexities are usually large,
(3) they cannot be built on a given arbitrary grammar. They use at least
$O(z\log(n/z)\log n)$ bits (which is an {\em upper bound} to our space
complexity) and can search as fast as in $O(m+\log^\epsilon z + 
occ(\log^\epsilon z + \log\log n))$ time for any constant 
$\epsilon>0$ \cite{CE18}, decreasing to $O(m+occ\log\log n)$ time with 
$O(z\log(g/z)\log\log z\log n)$ bits of space \cite{BEGV18}.
Gagie et al.~\cite{GGKNP12} can depart from any given grammar, but add
some extra space so that, within $O(G\log n + z\log\log z \log n)$ bits, they
can search in time $O(m^2+(m+occ)\log\log n)$. 

Recently, Navarro and Prezza
\cite{NP18} introduced a self-index of $O(\gamma\lg(n/\gamma)\lg n)$ bits, where
$\gamma \le z \le G$ is the size of any {\em attractor} of $T$ (which lower-%
bounds many other repetitiveness measures). They can search in time 
$O(m\log n + occ\log^\epsilon n)$. This result is theoretically appealing, but
again suffers from the drawbacks (1)--(3) above, and still its size dominates
only an upper bound on $G$.

In this article we introduce the first (as of the time of conference publication
\cite{CN12}, and still the only) grammar-based self-index that can be built from
{\em any given grammar} of size $G$, which uses $O(G\lg n)$ bits and whose 
search time depends only logarithmically on $n$, independently of the grammar
height. In addition, we give a practical implementation of the index and compare
it with different state-of-the-art indexes on repetitive collections,
showing that our index is also practical. In fact, the ability of our index to
build on any grammar has an important practical value, because it can be built
on top of compressors like RePair, which perform extremely well in practice.

The following theorem summarizes its properties; we note that the search 
time can be simplified to $O((m^2+occ)\lg G)$ because $\lg\lg n \le \lg G$.

\begin{theorem} \label{thm:main} 
Let a sequence $T[1..n]$ be represented by a
context-free grammar with $g$ symbols, size $G$ and height $h$. Then, for any
$0<\epsilon\le 1$, there exists a data structure using at most $G\lg n +2G\lg g
+ \epsilon\, g\lg g + o(G\lg g)+O(G)$ bits that finds the $occ$ occurrences of
any pattern $P[1..m]$ in $T$ in time $O((m^2/\epsilon)\lg\lg n +
(m+occ)(1/\epsilon+\lg g/\lg\lg g))$. It can extract any substring of
length $\ell$ from $T$ in time $O(\ell+h\lg(G/h))$. The structure can be built
in $O(n+G\lg G)$ time and $O(n\lg n)$ bits of working space. 
\end{theorem}

Note that the extraction time still depends on the grammar height. To improve 
it, we can include the structure of Belazzougui et al.~\cite{BPT15}, which
adds $O(G\lg n)$ bits. Within that space we derive a coarser
version of our result.

\begin{corollary} \label{cor:main} Let a sequence $T[1..n]$ over alphabet
$[1..\sigma]$ be represented by a context-free grammar with $g$ symbols and of
size $G$. Then there exists an index requiring $O(G \lg n)$ bits that
finds the $occ$ occurrences of any pattern $P[1..m]$ in $T$ in time 
$O(m^2 + (m+occ)\lg^\epsilon g)$, for any constant $\epsilon>0$,
and extracts any 
substring of length $\ell$ from $T$ in time $O(\ell/\lg_\sigma n+\lg n)$. 
\end{corollary}

In the rest of the article we describe our structure. First, we
preprocess the grammar to enforce several invariants useful to ensure our time
complexities. Then we use a data structure for binary relations
\cite{BCN13} to find the ``primary'' occurrences of $P$, that is, those formed
when concatenating symbols in the right hand side of a rule. To get rid of the
factor $h$ in this part of the search, we extend a technique \cite{GKPS05}
to extract
the first $m$ symbols of the expansion of any nonterminal in time $O(m)$. To
find the ``secondary'' occurrences (i.e., those that are found as the result
of the nonterminal containing primary occurrences being mentioned elsewhere),
we use a pruned representation of the parse tree of $T$. This tree is
traversed upwards for each secondary occurrence to report. The grammar
invariants introduced ensure that those traversals amortize to a constant
number of steps per occurrence reported. In this way we get rid of the factor
$h$ on the secondary occurrences too. 

We also show that our structure is practical. In Section~\ref{sec:exper} we
implement the index of Theorem~\ref{thm:main} and show that it outperforms
the preceding grammar-based index \cite{CN11}, even in its optimized form
\cite{CFMN16}, and it becomes a valid space/time tradeoff to the Lempel-Ziv
based self-index \cite{KN12} (also in optimized form \cite{CFMN16}). Our
expermental results show that, while the technique to speed up the extraction
\cite{GKPS05} does not have an impact in practice, the idea to amortize the 
cost of finding the secondary occurrences does speed up the index significantly
in practice.

The main differences with our conference version \cite{CN12} are improved
theoretical complexities, the whole implementation and experimental results, 
and an expanded and improved writing.

\section{Basic Concepts}

\subsection{Sequence Representations}
\label{sec:seqs}

Our data structures use succinct representations of sequences. Given a
sequence $S[1..N]$, over the alphabet $\Sigma$, we need to support the
following operations:

\begin{itemize}
\item
$\access(S,i)$ retrieves the symbol $S[i]$;
\item
$\rank_a(S,i)$ counts the number of occurrences of $a$ in $S[1..i]$;
\item
 $\select_a(S,j)$ computes the position where the $j$th $a$ appears in $S$.
\end{itemize}

For the case $|\Sigma| = 2$ (i.e., bitmaps), all the operations can be supported in
$N+o(N)$ bits and constant time \cite{Cla96}.
Raman et al.~\cite{RRR07} proposed two compressed
representations that are useful when the number $N_1$ of $1$s in $S$ is small.
One is the ``fully indexable dictionary'' (FID). It takes
$N_1\lg\frac{N}{N_1}+O(N_1)+o(N)$ bits of space and supports all the
operations in constant time. A weaker one is the ``indexable dictionary''
(ID), which takes $N_1\lg\frac{N}{N_1}+O(N_1+\lg\lg N)$ bits of space and supports
in constant time queries $\access(S,i)$, $\rank(S,i)$ if $S[i]=1$, and
$\select_1(S,j)$.

For general sequences, we will use a representation \cite{BN14} that requires
$N\lg |\Sigma| + o(N\lg |\Sigma|)$ bits and solves $\access(S,i)$ in $O(1)$ time and 
$\select(S,j)$ in any time in $\omega(1)$ (as a function of $|\Sigma|$), or vice 
versa; $\rank(S,i)$ takes time $O(\lg\lg_w |\Sigma|)$, on a RAM machine of $w$ bits.

\subsection{Labeled Binary Relations}
\label{sec:lbinrel}

A labeled binary relation is a binary relation $\Rel \subseteq
A\times B$, where $A=[1..n_1]$ and $B=[1..n_2]$, augmented with a function
$\Lab:A\times B \rightarrow L\cup\{\perp\}$, $L=[1..\ell]$, that defines
labels for each pair in $\Rel$, and $\perp$ for pairs that are not in $\Rel$.
Let us identify $A$ with the columns and $B$ with the rows in a table.
In our case, each element of $A$ will be associated with exactly one element 
of $B$, so $|\Rel|=n_1$. We augment a representation of unlabeled binary 
relations \cite{BCN13} with a plain string $S_\Lab[1..n_1]$ on alphabet
$[1..\ell]$, where $S_\Lab[i]$ is the label of the pair of column $i$.
The total space of this structure is  $n_1(\lg n_2 + \lg \ell)+o(n_1\lg n_2)$ 
bits. With this representation we can answer, among others, the following 
queries of interest in this article: 
\begin{enumerate}
\item Find the label of the element $b$ associated with a given $a$, 
$S_\Lab[a]$, in $O(1)$ time. 
\item Given $a_1, a_2, b_1$, and $b_2$, enumerate the $k$ pairs $(a,b)\in\Rel$ 
such that $a_1\leq a \leq a_2$ and $b_1\leq b \leq b_2$, in time
$O((k+1)(1+\lg n_2/\lg\lg(n_1+n_2)))$.
\end{enumerate}

\subsection{Succinct Tree Representations}
\label{sec:trees}

There are many representations for trees $\Tree$ with $N$ nodes that take
$2N+o(N)$ bits of space. In this paper we use one called Fully-Functional (FF)
\cite{NS14}, which in particular answers in constant time the following 
operations (node identifiers $v$ are associated with a position in $[1..2N]$): 
\begin{itemize}
\item $\node_{\Tree}(p)$ is the node with preorder number $p$; 
\item $\preorder_{\Tree}(v)$ is the preorder number of node $v$; 
\item $\leafrank_{\Tree}(v)$ is the number of leaves to the left of $v$; 
\item $\leafselect_{\Tree}(j)$ is the $j$th leaf;
\item $\intrank_{\Tree}(v)$ is the number of internal nodes before $v$, in preorder; 
\item $\intselect_{\Tree}(j)$ is the $j$th internal node, in preorder;
\item $\numleaves_{\Tree}(v)$ is the number of leaves below $v$; 
\item $\parent_{\Tree}(v)$ is the parent of $v$; 
\item $\child_{\Tree}(v,k)$ is the $k$th child of $v$; 
\item $\nextsibling_{\Tree}(v)$ is the next sibling of $v$; 
\item $\degree_{\Tree}(v)$ is the number of children of $v$;
\item $\depth_{\Tree}(v)$ is the depth of $v$; and 
\item $\levelanc_{\Tree}(v,k)$ is the $k$th ancestor of $v$. 
\end{itemize}

The FF representation is obtained by traversing the tree in DFS order and
appending to a bitmap a $1$ when we arrive at a node, and a $0$ when we leave
it. The operations $\leafrank$, $\leafselect$, $\intrank$, and $\intselect$ are
not discussed so widely in the literature. In the FF sequence $F[1..2N]$, each 
internal node starts with a bit $1$ followed by another $1$, and each leaf is 
represented by a $1$ followed by a $0$. The same mechanisms described in 
Section~\ref{sec:seqs} to support $\rank$ and $\select$ for $0$s and $1$s on 
bitmaps are easily extended to support two-bit operations, within $o(N)$ 
extra bits. Therefore, we implement
$\leafrank(i)=\rank_{10}(F,i-1)$, $\leafselect(j)=\select_{10}(F,j)$,
$\intrank(i)=\rank_{11}(F,i-1)$, and $\intselect(j)=\select_{11}(F,j)$, all
in constant time.

\section{Preprocessing the Grammar}
\label{sec:preproc}

We will work on a given context-free grammar $\G$ that generates a
single string $T[1..n]$ over alphabet $\Sigma=[1..\sigma]$, formed by 
$g$ (terminal and nonterminal) symbols. The
$\sigma \le g$ terminal symbols come from an alphabet $\Sigma=[1..\sigma]$,%
\footnote{Non-contiguous alphabets can be handled with an ID 
(Section~\ref{sec:seqs}) that marks the symbols present in $T$.} and
then $\G$ contains $g-\sigma$ rules of the form $X_i \rightarrow \alpha_i$, 
exactly one per nonterminal. The sequence $\alpha_i$, called 
the {\em right-hand side} of the rule, is the sequence of terminal and 
non-terminal symbols generated by $X_i$ (without recursively unrolling rules). 
We call $G = \sum |\alpha_i|$ the {\em size} of $\G$. Note it holds $\sigma \le
G$, since the terminals must appear in the right-hand sides.  We assume all the
nonterminals are used to generate the string; otherwise unused rules can be
found and dropped in $O(G)$ time. 
The grammar cannot have loops since it generates a finite string $T$. 

Let $X_s$ be always the start symbol (despite of successive symbol renamings).
We call $\F(X_i)$ the single string generated by $X_i$, that is $\F(a)=a$ 
for terminals $a$ and $\F(X_i) = \F(X_{i_1})\cdots \F(X_{i_k})$ for nonterminals
$X_i \rightarrow X_{i_1} \ldots X_{i_k}$. The grammar $\G$ generates the text 
$T = \mathcal{L}(\G)=\F(X_s)$.

For the purpose of building our index, we preprocess $\G$ as follows.  

\begin{itemize}
\item 
First, for each terminal symbol $a\in\Sigma$
present in $\G$ we create a rule $X_a \rightarrow a$, and replace all other
occurrences of $a$ in the grammar by $X_a$. As a result, the grammar contains
exactly $g$ nonterminal symbols $\X = \{X_1, \ldots, X_g\}$, each associated
with a rule $X_i \rightarrow \alpha_i$, where $\alpha_i \in \Sigma$ or
$\alpha_i$ is a sequence of elements in $\X$. 
\item
Any rule that generates just one single nonterminal $X_i \rightarrow X_j$, or
the empty string, $X_i \rightarrow \varepsilon$,
is removed by replacing $X_i$ by $X_j$ or by $\varepsilon$ everywhere. This
decreases $g$ without increasing $G$.
\item
We further preprocess $\G$ to enforce the property that any nonterminal $X_i$,
except $X_s$ and those $X_a \rightarrow a \in \Sigma$, must be mentioned in at
least two right-hand sides. We traverse the rules of the grammar, count the
occurrences of each symbol, and then rewrite the rules, so that only the rules
of those $X_i$ appearing more than once (or the excepted symbols) are preserved,
and as we rewrite their right-hand sides, we replace any (non-excepted) $X_i$ that
appears once by its right-hand side $\alpha_i$. This transformation takes $O(G)$
time and can only reduce $G$ and $g$.
%
%
%
%
%

\item
Our last preprocessing step, and the most expensive one, is to renumber the
nonterminals so that $i<j \Leftrightarrow \F(X_i)^{rev}<\F(X_j)^{rev}$, where
$S^{rev}$ is string $S$ read backwards (the purpose of this renumbering
will be apparent
later).  The sorting can be done in time $O(n+g\lg g)$ and $O(n\lg n)$ bits of
space, following the same approach as in previous work \cite{CN11}. This
process dominates the construction time. 
\end{itemize}

From now on, $g$ will refer to the number of rules in the transformed grammar 
$\G$ (i.e., the number of terminal and nonterminal symbols in the original
grammar, minus possible reductions). Instead, 
$G$ will still be the size of the original grammar (the transformed one has 
size at most $G+\sigma$).

\section{Main Index Structure}

We define a structure that will be key in our index.

\begin{definition} The {\em grammar tree} of $\G$ is a general tree $\Tree_\G$
with nodes labeled in $\X$. Its root is labeled $X_s$ and its topology is
obtained by pruning the parse tree of $T$ with two rules: (1) in a
left-to-right DFS traversal, in each noded except the first time a nonterminal 
$X_i$ is found, its subtree is pruned and the node becomes a leaf; (2) whenever
$X_a \rightarrow a$ is found, it is pruned too, leaving $X_a$ as a leaf. 
We say that each $X_i$ is {\em defined} in the only internal node of 
$\Tree_\G$ labeled $X_i$.
\end{definition}

\begin{figure}[t!]
\begin{center}
\includegraphics[width=0.8\textwidth]{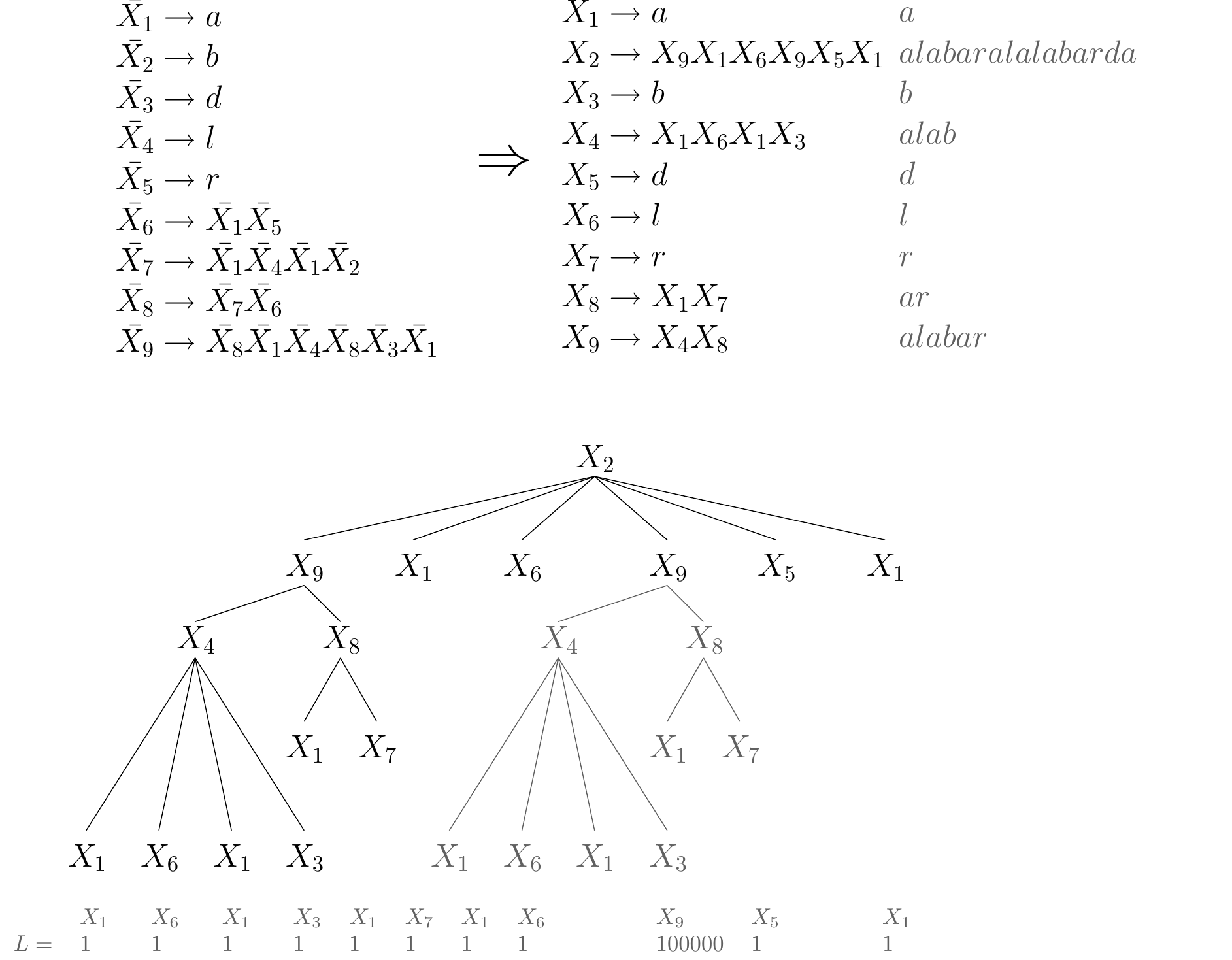}
\caption{At the top left, a grammar $\G$ generating string
{\tt "alabaralalabarda"}. At the top right, our reordering of the grammar and
strings $\F(X_i)$. On the bottom, the grammar tree $\Tree_\G$ in black; the
whole parse tree includes also the grayed part. Below the tree we show our
bitmap $L$ (Section~\ref{sec:extractarbitrary}).}
\label{fig:example_grammar}
\end{center}
\end{figure}

Since each right-hand side $\alpha_i \not\in\Sigma$ is written once in the
tree as the children of $X_i$, and the root $X_s$ is written once, the total 
number of nodes in $\Tree_\G$ is $G+1$.
The number of internal nodes is $g-\sigma$, and the number of leaves is
$G+1-g+\sigma$.
Figure \ref{fig:example_grammar} shows the reordering and grammar tree for a
grammar generating the string {\tt "alabaralalabarda"}.

The grammar tree partitions $T$ in a way that is useful for finding
occurrences, using a concept that dates back to K\"arkk\"ainen \cite{Kar99}.

\begin{definition} Let $X_{l_1}, X_{l_2},\ldots$ be the nonterminals labeling
the consecutive leaves of $\Tree_\G$. Let $T_i = \F(X_{l_i})$, then $T = T_1 T_2
\ldots$ is a partition of $T$ according to the leaves of $\Tree_\G$. We say that
an occurrence of a pattern $P$ is {\em primary} relatively to the given 
partition if it spans more than one $T_i$. The other occurrences are called {\em
secondary}. \end{definition}

Our self-index will represent $\G$ using two main components. One represents the
grammar tree $\Tree_\G$ using an FF representation (Section~\ref{sec:trees})
and a sequence of labels (Section~\ref{sec:seqs}).
This will be used to extract the text and decompress
rules. When augmented with a secondary trie $\Tree_S$ storing leftmost/rightmost
paths in $\Tree_\G$, the representation will expand any prefix/suffix of a rule
in optimal time~\cite{GKPS05}.

The second component in our self-index corresponds to a labeled binary relation
(Section~\ref{sec:lbinrel}),
where $B=\X$ and $A$ is the set of proper suffixes starting at positions $j+1$
of rules $\alpha_i$: $(\alpha_i[j],\alpha_i[j+1 ..])$ will be related for all
$X_i \rightarrow \alpha_i$ and $1 \le j < |\alpha_i|$. The labels are numbers in
the range $[1..G+1]$; we specify their meaning later. This binary relation will
be used to find the primary occurrences of the search pattern. Secondary
occurrences will be tracked in the grammar tree.

\section{Extracting Text}
\label{sec:extract}

We first describe a simple structure that extracts the prefix of length $\ell$
of any rule in $O(\ell+h)$ time.  We then augment this structure to support
extracting any substring of length $\ell$ in time $O(\ell+h\lg(G/h))$, and
finally augment it further to retrieve the prefix or suffix of length $\ell$
of any rule in
optimal $O(\ell)$ time. This last result is fundamental for supporting
searches, and is obtained by extending the structure proposed by Gasieniec et
al.~\cite{GKPS05} for SLPs to general context-free grammars.
The improvement does not work for extracting arbitrary substrings, as
in that case one has to find first the nonterminals that must be expanded.
This subproblem is not easy to solve, especially within little space 
\cite{BLRSRW15}.

As said, we represent the topology of the grammar tree $\Tree_\G$ using FF
(Section~\ref{sec:trees}), using $O(G)$ bits. 
The sequence of labels associated with the tree nodes is stored in preorder in 
a sequence $X[1..G+1]$ using $G\log g + o(G\log g)$ bits with the representation
described in Section~\ref{sec:seqs}, where we choose constant time for 
$\access(X,i)=X[i]$ and $O(\lg\lg_w G)$ time for $\select_a(X,j)$.

We also store a bitmap $Y[1..g]$ that marks the rules of the form
$X_i\rightarrow a \in \Sigma$ with $1$s. Since the rules have been renumbered
in (reverse) lexicographic order, every time we find a rule $X_i$ such that
$Y[i]=1$, we can determine the terminal symbol it represents as $a=\rank_1(Y,i)$
in constant time. 
In our example of
Figure~\ref{fig:example_grammar} this bitmap is $Y=101011100$.

\subsection{Expanding Prefixes of Rules}
\label{sec:expandprefix}

Expanding a rule $X_i$ that does not correspond to a terminal is done as
follows. By the definition of $\Tree_\G$, the first left-to-right occurrence of
$X_i$ in sequence $X$ corresponds to the definition of $X_i$; all the others are
leaves in $\Tree_\G$. Therefore, $v=\node_{\Tree_\G}(\select_{X_i}(X,1))$ is the
node in $\Tree_\G$ where $X_i$ is defined. We then traverse the subtree rooted 
at $v$ in DFS order. Every time we reach a leaf $u$, we compute its label
$X_j=X[\preorder_{\Tree_\G}(u)]$, and either output a terminal if $Y[j]=1$ or
recursively expand $X_j$. This is in fact a traversal of the {\em parse tree}
starting at node $v$, using the grammar tree instead. The traversal to extract
the first $\ell$ terminals takes $O(\ell+h_v)$ steps, where $h_v \le h$ is the 
height of the parsing subtree rooted at $v$. In particular, if we extract the 
whole sequence $\F(X_i)$, we perform $O(\ell)=O(\F(X_i))$ steps, since we have 
removed unary paths in the preprocessing of $\G$ and thus $v$ has $\F(X_i) > 
h_v$ leaves in the parse tree. The only obstacle to
having constant-time steps are the queries $\select_{X_i}(X,1)$. As these are
only for the position 1, they correspond to finding the internal node of 
$\Tree_\G$ that defines $X_i$. We then store a permutation $\pi[1..g-\sigma]$ 
so that $X_i$ is defined at the node $\intselect_{\Tree_\G}(\pi[\rank_0(Y,i)])$,
which is computed in constant time using $g\lg g$ bits for $\pi$.

The total space required for $\Tree_\G$, considering the FF representation,
sequence $X$, bitmap $Y$, and permutation $\pi$, is $G\lg g + g\lg g + 
o(G\lg g)+O(G)$ bits.\footnote{It could be that $g=O(1)$, so $o(G\lg g)$ does
not necessarily absorb $O(G)$.} We reduce the space 
to $G\lg g + \epsilon\,g\lg g+o(G\lg g) +O(G)$,
for any $0<\epsilon\le 1$, by removing some redundancy: We form a reduced
sequence $X'[1..G-g+\sigma+1]$ where the labels of the internal nodes are 
removed. We can still access any $X[i] = X'[\leafrank_{\Tree_\G}(v)+1]$, with $v=\node_{\Tree_\G}(i)$,
if $v$ is a leaf. If $v$ is an internal node, we have
$X[i] = \select_0(Y,\pi^{-1}[\intrank_{\Tree_\G}(v)+1])$. 
We can also support general $\select$ on $X$:
$\select_{X_i}(X,j) = \preorder_{\Tree_\G}(\leafselect_{\Tree_\G}(\select_{X_i}(X',j-1)))$ for $j>1$.

Thus, we can use $X'$ instead of $X$, at the cost of having to compute $\pi^{-1}$.
To do this, we use the representation of Munro et al.~\cite{MRRR12}
that takes $(1+\epsilon)g\lg g$ bits and computes any $\pi[i]$ in constant time
and any $\pi^{-1}[j]$ in time $O(1/\epsilon)$. This yields the promised space.
The time to access 
$X[i]$ is now $O(1/\epsilon)$. Although this will have an impact later, we 
note that for extraction we only access $X$ at leaf nodes, where it 
takes constant time.

\subsection{Extracting Arbitrary Substrings}
\label{sec:extractarbitrary}

In order to extract any given substring of $T$, we add a bitmap $L[1..n]$
that marks with a $1$ the first position of each $T_i$ in $T$
(see Figure~\ref{fig:example_grammar}).
We can then compute the starting position of any node $v \in \Tree_\G$ as
$p(v) = \select_1(L,\leafrank_{\Tree_\G}(v)+1)$.

To extract $T[p..p+\ell-1]$, we binary search the children of the root of
$\Tree_\G$, to find the child $u$ covering position $p$. If $u$ is a leaf
representing a nonterminal, we go to its definition 
$v \in \Tree_\G$, translate position $p$ to the area below the new node $v$
(i.e., $p$ becomes $p-p(u)+p(v)$), and continue recursively from $v$. At some 
point we reach the terminal node $X_i \rightarrow a$ covering position $p$, 
and from there on we extract the symbols rightwards. Just as
before, the total number of steps is $O(\ell+h)$. Yet, the $h$ steps require
binary searches. As there are at most $h$ binary searches among the children
of different tree nodes, and there are $G+1$ nodes, at worst the binary
searches cost $O(h \lg (G/h))$. The total cost is $O(\ell + h \lg (G/h))$.

The number of $1$s in $L$ is at most $G$. Since we only need $\select_1$ on
$L$, we can use an ID representation (Section~\ref{sec:seqs}), requiring
$G\lg(n/G)+O(G+\lg\lg n) = G\lg(n/G)+O(G)$ bits (since $G \ge \lg n$ in any
grammar). The total space then becomes $G\lg g + G\lg(n/G) + \epsilon\, g\lg g +
o(G\lg g)+O(G)$ bits.

Instead,
if we implement $L$ with an Elias-Fano structure \cite{Eli74,Fan71}, and
augment each sub-universe of size $n/G$ with a sampled predecessor data 
structure, we use $G\lg(n/G)+o(G\lg n)$ bits for $L$ and can solve $rank_1$
queries on $L$ in time $O(\lg\lg(n/G))$ (cf.\ \cite[Sec.~4.2]{BN14}). 
Thus, instead of doing successive
binary searches in the path towards the leaf covering $p$, we compute the
area of that leaf directly with $1+rank_1(L,p)$. Therefore the total time
becomes $O(h\lg\lg(n/G))$, because there can still be $h$ jumps to other
parts of the tree. We omit this results for simplicity.

\subsection{Optimal Expansion of Rule Prefixes and Suffixes}
\label{sec:extractoptimal}

Our improved version builds on the proposal by Gasieniec et al.~\cite{GKPS05}.
We extend their representation to handle
general grammars instead of only SLPs. Using their notation, call $S(X_i)$ the
string of labels of the nodes in the path from any node labeled $X_i$ to its
leftmost leaf in {\em the parse tree} (we take as leaves the nonterminals $X_a
\in \X$ with $X_a \rightarrow a$, not the terminals $a\in\Sigma$). We insert 
all the strings $S(X_i)^{rev}$ into a trie $\Tree_S$. Note that each symbol 
$X_i$ appears only once in
$\Tree_S$ \cite{GKPS05}, thus $\Tree_S$ has $g$ nodes. Again, we represent the 
topology
of $\Tree_S$ using FF. Its sequence of labels $X_S[1..g]$ turns out to
be a permutation of $[1..g]$. We represent it once again with the structure
\cite{MRRR12} that takes $(1+\epsilon)g\lg g$ bits and computes any $X_S[i]$ in
constant time and any $X_S^{-1}[j]$ in time $O(1/\epsilon)$.

We can determine the first terminal in the expansion of $X_i$, which labels
node $v \in \Tree_S$, as follows. Since the last symbol in $S(X_i)$ is a
nonterminal $X_a$ with $X_a \rightarrow a$ for some $a \in \Sigma$, it follows 
that $X_i$
descends in $\Tree_S$ from $X_a$, which is a child of the root.  This node is
$v_a = \levelanc_{\Tree_S}(v,\depth_{\Tree_S}(v)-1)$. Then $a = \rank_1(Y,X_S[\preorder_{\Tree_S}(v_a)])$.

\begin{figure}[t]
\begin{center}
\includegraphics[width=9cm]{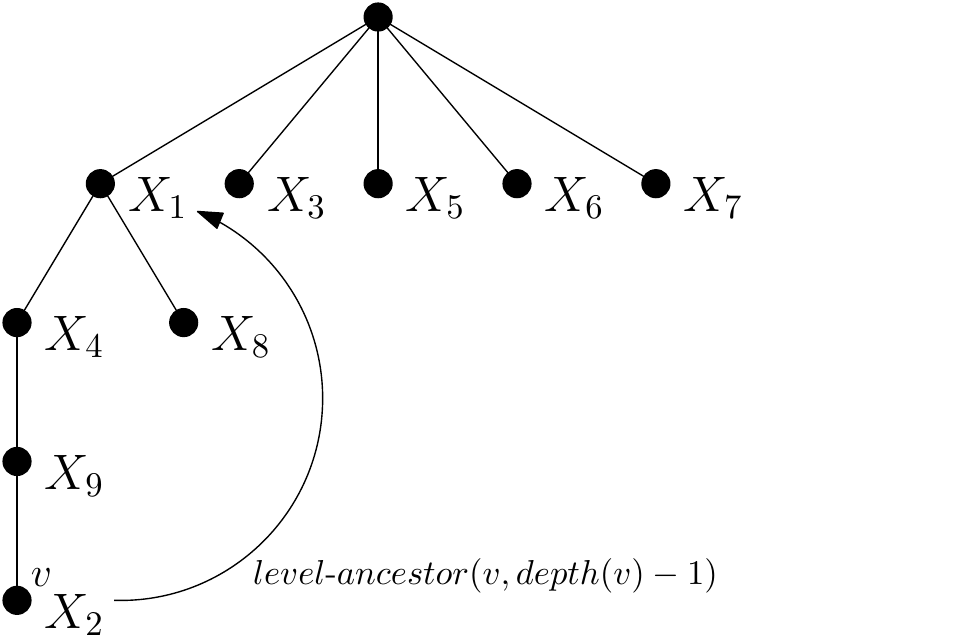}
\caption[Leftmost path trie for extracting rules]{Example of the trie of leftmost paths for the grammar of
Figure~\ref{fig:example_grammar}. The arrow pointing from $X_2$ to
$X_1$ illustrates the procedure to determine the first terminal symbol
generated by $X_2$.}
\label{fig:example_trie}
\end{center}
\end{figure}

Figure \ref{fig:example_trie} shows an
example of this query in the trie for the grammar presented
in Figure \ref{fig:example_grammar}.

A prefix of $X_i$ is then extracted as follows. First, we obtain the 
corresponding
node $v \in \Tree_S$ as $v = \node_{\Tree_S}(X_S^{-1}[X_i])$. Then we obtain the leftmost
symbol of $v$ as explained. The remaining symbols descend from the second and
following children, in the parse tree, of the nodes in the upward path from a
node labeled $X_i$ to its leftmost leaf, or which is the same, of the nodes in
the downward path from the root of $\Tree_S$ to $v$. Therefore, for each node
$u$ in the list $\levelanc_{\Tree_S}(v,\depth_{\Tree_S}(v)-2),
\ldots,\parent_{\Tree_S}(v),v$, we map $u$ to $x \in \Tree_\G$,
$x = \node_{\Tree_\G}(\select_{X_j}(X,1))$ where $X_j =
X_S[\preorder_{\Tree_S}(u)]$. Once $x$ is found, we recursively expand its
children, from the second onwards, by mapping them back to $\Tree_S$. Charging the cost to the new symbol to be expanded, and since there are no
unary paths, it follows that we carry out $O(\ell)$ steps to extract the first
$\ell$ symbols, and the extraction is real-time \cite{GKPS05}.  All costs per
step are $O(1)$ except for the $O(1/\epsilon)$ to access $X_S^{-1}$.

For extracting suffixes of rules in $\G$, we need another version of $\Tree_S$
that stores the rightmost paths. This yields the following result.

\begin{lemma} \label{lem:extract} Let a sequence $T[1..n]$ be represented by a
context-free grammar with $g$ symbols, size $G$, and height $h$. Then, for any
$0<\epsilon \le 1$, there exists a data structure using at most $G\lg g +
G\lg(n/G) + (2+\epsilon)g\lg g + o(G\lg g)+O(G)$ bits of space that extracts any
substring of length $\ell$ from $T$ in time $O(\ell+h\lg(G/h))$, and a prefix
or a suffix of length $\ell$ of the expansion of any nonterminal in time
$O(\ell/\epsilon)$. \end{lemma}


\section{Locating Patterns}
\label{sec:search}

A secondary occurrence of the pattern $P$ inside a leaf of $\Tree_\G$ labeled
by a symbol $X_i$ occurs as well in the internal node of $\Tree_G$ where $X_i$
is defined. If that occurrence is also secondary, then it occurs inside a
child $X_j$ of $X_i$, and we can repeat the argument with $X_j$ until finding
a primary occurrence inside some $X_k$. Thus, to find all the secondary
occurrences, we can first spot the primary occurrences, and then find all the
copies of the nonterminals $X_k$ that contain the primary occurrences, as well
as all the copies of the nonterminals that contain $X_k$, recursively.

As before, we base our approach on the strategy proposed by
K{\"a}rkk{\"a}inen \cite{Kar99} to find the primary occurrences of
$P=p_1p_2\ldots p_m$. K{\"a}rkk{\"a}inen considers the $m-1$ partitions $P = P_1
\cdot P_2$, $P_1 = p_1 \ldots p_i$ and $P_2 = p_{i+1}\ldots p_m$, for $1 \le i <
m$. In our case, for each partition we will find all the nonterminals $X_k
\rightarrow X_{k_1} X_{k_2} \ldots X_{k_r}$ such that $P_1$ is a suffix of some
$\F(X_{k_i})$ and $P_2$ is a prefix of $\F(X_{k_{i+1}})\ldots \F(X_{k_r})$. This
finds each primary occurrence exactly once. The secondary occurrences are then
tracked in the grammar tree $\Tree_\G$. We handle the case $m=1$ by finding
all occurrences of $X_j$, where $X_j \rightarrow p_1$, in $\Tree_\G$ using $\select_{X_j}$ over the 
sequence of labels, and treating them as primary occurrences.

\subsection{Finding Primary Occurrences}

As anticipated at the end of Section~\ref{sec:preproc}, we store a binary
relation $\Rel \subseteq A \times B$ to find the primary occurrences. It has
$g$ rows labeled $X_i$, for all $X_i \in \X = B$, and $G-g$ columns.
Each column corresponds
to a distinct proper suffix $\alpha_i[j+1..]$ of a right-hand side $\alpha_i$.
The labels belong to $[1..G+1]$. The relation contains one pair per column:
$(\alpha_i[j], \alpha_i[j+1..]) \in \Rel$ for all $1 \le i \le g$ and $1 \le j
< |\alpha_i|$. Its label is the preorder of the $(j+1)$th child of the node
that defines $X_i$ in $\Tree_\G$. The space for the binary relation is
$(G-g)(\lg g+\lg G)+o(G\lg g)$ bits.

Recall that, in our preprocessing, we have sorted $\X$ according to the
lexicographic order of $\F(X_i)^{rev}$. We also sort the suffixes $\alpha_i[j+1
..]$ lexicographically with respect to their expansion, that is
$\F(\alpha_i[j+1])\F(\alpha_i[j+2]) \ldots \F(\alpha_i[|\alpha_i|])$. This can
be done in $O(n+G\lg G)$ time in a way similar to how $\X$ was sorted: Each
suffix $\alpha_i[j+1 ..]$, labeled $p$, can be associated with the substring
$T[\select_1(L,\leafrank_{\Tree_\G}(\node_{\Tree_\G}(p))+1) ..
\select_1(L,\leafrank_{\Tree_\G}(v) +
\numleaves_{\Tree_\G}(v)+1)-1]$, where $v$ is the parent of
$\node_{\Tree_\G}(p)$. Then we can proceed as in previous constructions for 
SLPs \cite{CN11}.

\begin{figure}[t]
\begin{center}
\includegraphics[width=5.5cm]{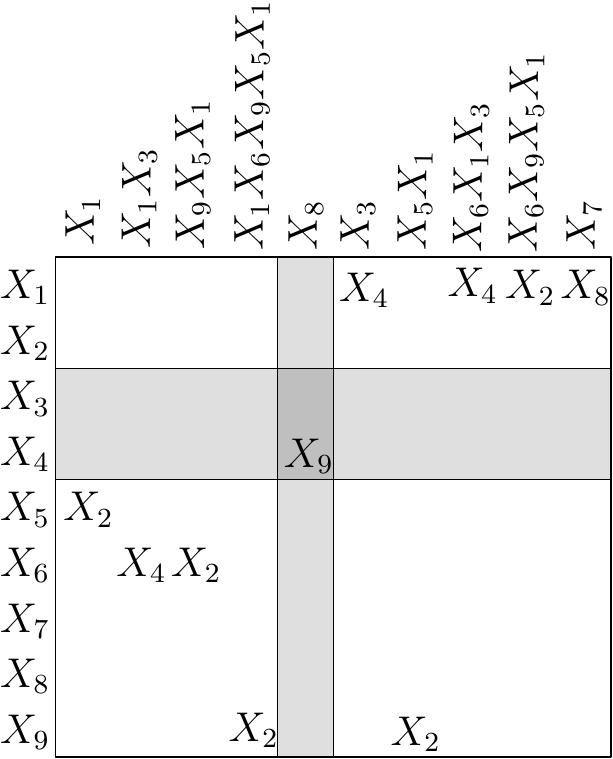}
\caption[Example of search on general grammars]{Relation $\Rel$ for the grammar presented in Figure
\ref{fig:example_grammar}. The highlighted ranges correspond to the
result of searching for $b\cdot ar$, where the single primary occurrence
corresponds to $X_9$.}
\label{fig:binrel}
\end{center}
\end{figure}

Figure \ref{fig:binrel} illustrates how
$\Rel$ is used for the grammar presented in Figure
\ref{fig:example_grammar}.

Given $P_1$ and $P_2$, we first find the range of rows whose expansions finish
with $P_1$, by binary searching for $P_1^{rev}$ in the expansions
$\F(X_i)^{rev}$. Each comparison in the binary search needs to extract $|P_1|$
terminals from the suffix of $\F(X_i)$. According to Lemma~\ref{lem:extract},
this takes $O(|P_1|/\epsilon)$ time. Similarly, we binary search for the range
of columns whose expansions start with $P_2$. Each comparison needs to extract
$\ell=|P_2|$ terminals from the prefix of
$\F(\alpha_i[j+1])\F(\alpha_i[j+2])\ldots$. Let $r$ be the column we wish to
compare to $P_2$. We extract the label $p$ associated with the column in constant
time. Then we extract the first $\ell$ symbols from the expansion of
$node_{\Tree_\G}(p)$. If $node_{\Tree_\G}(p)$ does not have enough symbols, we
continue with $\mathit{nextsibling}_{\Tree_\G}(p)$, and so on, until we extract
$\ell$ symbols or we exhaust the suffix of the rule. According to
Lemma~\ref{lem:extract}, this requires time $O(|P_2|/\epsilon)$. Thus our two
binary searches require time $O((m/\epsilon)\lg G)$.

This time can be further improved by building a trie of sampled expansions. We 
sample expanded strings at regular intervals and store them in a Patricia tree 
\cite{Mor68}.  We first search for the pattern in the
Patricia tree, and then complete the process with a binary search between two
sampled strings (we first verify the correctness of the Patricia search by
checking that our pattern is actually within the range found).  By sampling
one out of $\lg n$ strings, the search time becomes 
$O((m/\epsilon)\lg\lg n)$ and we only require $O(G)$ bits of extra space, since
the Patricia tree needs $O(\lg n)$ bits per node.\footnote{We could push it
a bit further, for example sampling one out of $\lg n \lg\lg g / \lg g$
strings to obtain $o(G\lg g)+O(G)$ bits of extra space and a search time
of $O\left((m/\epsilon)\lg\left(\frac{\lg n \lg\lg g}{\lg g} \right)\right)$,
but we opt for a simpler formula.}

Once we identify a range of rows $[a_1,a_2]$ and of columns $[b_1,b_2]$, we
retrieve all the $k$ points in the rectangle and their labels in time
$O((k+1)(1+\lg g / \lg\lg G))$. The parents of all the
nodes $node_{\Tree_\G}(p)$, for each point $p$ in the range, correspond to
the primary occurrences. In Section~\ref{sec:tracking} we show how to report
primary and secondary occurrences starting directly from those positions
$node_{\Tree_\G}(p)$.

We have to carry out this search for $m-1$ partitions of $P$, whereas each
primary occurrence is found exactly once. Calling $occ$ the number of primary
occurrences, the total cost of this part of the search is
$O((m^2/\epsilon)\lg\lg n + (m+occ) (1+\lg g / \lg\lg G))$.

\subsection{Tracking Secondary Occurrences through the Grammar Tree}
\label{sec:tracking}

The remaining problem is how to track all the secondary occurrences triggered
by a primary occurrence, and how to report the positions where these occur in
$T$. Given a primary occurrence for partition $P=P_1\cdot P_2$ located at
$v=\node_{\Tree_\G}(p)$, we obtain the starting position of $P$ in $T$ by
moving towards the root while keeping count of the offset between the
beginning of the current node and the occurrence of $P$. Initially, for node
$v$ itself, this is $l = -|P_1|$. Now, while $v$ is not the root, we set $l
\leftarrow l + \select_1(L,\leafrank_{\Tree_\G}(v)+1) -
\select_1(L,\leafrank_{\Tree_\G}(\parent_{\Tree_\G}(v))+1)$, and $v
\leftarrow \parent_{\Tree_\G}(v)$. When we reach the root, the occurrence of
$P$ starts at $l$.

It seems that we are doing this $h$ times in the worst case, since we need to
track the occurrence up to the root. In fact we might do so for some symbols,
but the total cost is amortized. Every time we move from $v$ to
$u=\parent_{\Tree_\G}(v)$, we know that $X[u]$ appears at least once more in
the tree. This is because our preprocessing (Section ~\ref{sec:preproc})
forces rules to appear at least twice or be removed. Thus $u$ defines
$X[u]$, but there are one or more leaves labeled $X[u]$, and we have to report
the occurrences of $P$ inside them all. For this sake we carry out
$\select_{X[u]}(X,i)$ for $i=2,3\ldots$ until spotting all those occurrences
(where $P$ occurs with the current offset $l$). We recursively track them to
the root of $\Tree_\G$ to find their absolute position in $T$, and recursively
find the other occurrences of all their ancestor nodes. The overall cost
amortizes to $O(1)$ steps per occurrence reported, as we can charge the cost
of moving from $v$ to $u$ to the other occurrence of $u$. If we report $occ$
secondary occurrences we carry out $O(occ)$ steps, each costing $O(\lg\lg g)$
time. 


\section{The Resulting Index}
\label{sec:res}

By adding up the space of Lemma~\ref{lem:extract} with that of the labeled
binary relation, and adding up the costs, we have our central result,
Theorem~\ref{thm:main}, where for simplicity we have replaced the cost
per occurrence of $1/\epsilon + \log\log g + \log g/\log\log G$ by just
$1/\epsilon + \log g / \log\log g$.

By using $\epsilon=\Theta(1)$ and $\epsilon=1/\lg\lg n$, we obtain two simpler
results.

\begin{corollary} \label{cor:main1} 
Let a sequence $T[1..n]$ be represented by a
context-free grammar with $g$ symbols and size $G$. Then, for any
constant $0<\epsilon\le 1$, there exists a data structure using at most 
$G\lg n +2G\lg g + \epsilon\, g\lg g + o(G\lg g)+O(G)$ bits that finds the 
$occ$ occurrences of any pattern $P[1..m]$ in $T$ in time $O(m^2\lg\lg n 
+ (m+occ)\lg g/\lg\lg g)$.
\end{corollary}

\begin{corollary} \label{cor:main2} 
Let a sequence $T[1..n]$ be represented by a
context-free grammar with $g$ symbols and size $G$. Then, 
there exists a data structure using at most 
$G\lg n +2G\lg g + o(G\lg g)+O(G)$ bits that finds the 
$occ$ occurrences of any pattern $P[1..m]$ in $T$ in time $O((m\lg\lg n)^2 
+ (m+occ)\lg g/\lg\lg g)$.
\end{corollary}

Finally, by using a larger geometric structure \cite{CLP11} for the binary
relation, and letting other structures use $O(G\lg n)$ bits, we obtain a 
somewhat faster structure, Corollary~\ref{cor:main}.

\section{Implementation and Experiments} \label{sec:exper}

\subsection{Implementation}

We implemented our grammar-based self-index on top of the library SDSL (\textit{Succinct Data Structures Library})\footnote{\texttt{https://github.com/simongog/sdsl-lite}}, which is implemented in C++11 and contains efficient implementations of several succinct data structures. 

To generate the grammar we use the RePair algorithm \cite{LM00}, in particular
Navarro's implementation\footnote{\texttt{http://www.dcc.uchile.cl/gnavarro/software/repair.tgz}}. RePair produces a binary grammar (i.e., all the rules have
2 symbols in their right-hand side) plus a long initial rule. 
We then postprocess the resulting grammar as required for our index, see
Section~\ref{sec:preproc}.

In repetitive collections it holds that $g \le G \ll n$; we also expect that
$\sigma \ll g$ for large texts. It follows that bitmaps $Y$ (of length $g$ and with $\sigma$ 1s) and  $L$ (of length $n$ and with less than $G$ 1s) are sparse. We then represent them using the class \texttt{sd\_vector} from SDSL. 

In the compressed grammar representation we use a permutation $\pi$ to find the node that defines a rule. We use the class \texttt{inv\_permutation\_support<$t$>} of SDSL, which gives access to the inverse permutation in at most $t$ steps, and fix $t$ to 32. We also use this structure for representing the labels of the tries $T_S$ in the optimal prefix/suffix extraction.

To support operations on the sequence of non-primary nodes, $X'$ is represented using the structure of Golynski et al.~\cite{GMR06} (\texttt{wt\_gmr} in
SDSL), because its alphabet is large and the sequence is almost incompressible.

Our grid representation is the same structure used in the
implementation of Claude and Navarro \cite{CN11} for labeled binary relations.

The topology of the grammar tree and of the sampled Patricia tries is 
represented with a variant of balanced FF (Section~\ref{sec:trees}) called 
DFUDS \cite{BDM+05}, which is faster in practice for moving towards children. 
The trees $T_S$, instead, are represented using FF \cite{NS14}, which is
more efficient for level ancestor queries. Both are implemented over the
parentheses support of SDSL (\texttt{bp\_support\_sada}).

We test four versions of our index, called {\bf\textsf{g-index}} in the experiments.
The variants whose name continue with \textsf{binary\_search} use plain binary
search on the rules prefixes/suffixes in order to find the row and column
intervals
on the grid. The variants whose name instead continue with \textsf{patricia\_tree} speed
up this process using a sampled Patricia tree, which takes one string every
4, 8, 16, 32, and 64 positions. On the other hand, the variants suffixed
\textsf{trie} use the structure of Gasieniec et al.~\cite{GKPS05}
(Section~\ref{sec:extractoptimal}) to extract rule prefixes/suffixes in optimal
time, whereas the variants suffixed \textsf{notrie} omit this structure and
extract the text from the rules in recursive form. Finally, the term
\textsf{gram} indicates that we add a short $q$-gram ($q=2,4,6,8,10,12$) with 
the prefix and suffix of the expansion of each nonterminal \cite{CFMN16}, to 
speed up extraction during binary searches. The strings are stored in a
dictionary compressed with Huffman and Front Coding. Since the $q$-grams are
limited, the binary search must be completed, either using plain decompression
of nonterminals (suffix \textsf{dfs}), the real-time prefix extraction (suffix
\textsf{trie}), or plain decompression speeded up by the same $q$-grams of the
nonterminals we find in the way (suffix \textsf{smp}).

Our implementation is available at
\texttt{https://github.com/apachecom/grammar\_improved\_index/}.

\subsection{Experimental Setup}

The experimental evaluation was carried out using the environment provided in {\em Pizza\&Chili} (\texttt{http://pizzachili.dcc.uchile.cl}). We compared our implementation with the available indexes in the state of the art that are most faithful with respect to different compressibility 
measures: 

\begin{description}
\item[\textsf{slp-index}\!\!]\footnote{\texttt{https://github.com/migumar2/uiHRDC/tree/master/uiHRDC/self-indexes/SLP}} is the only previous implementation of a
grammar-based index \cite{CN11}, using $O(G\log n)$ bits like ours. It does
not guarantee, however, logarithmic locating time per occurrence. It uses the
same RePair algorithm we use to build the index (a construction over the 
heuristically balanced version of RePair is called \textsf{slp-index-bal}). 
In its optimized version \cite{CFMN16}, it speeds up the binary searches by 
storing the $q$-gram prefixes of the strings expanded by each nonterminal,
as we use in the \textsf{gram} variant of \textsf{q-index}, yet here the best
values are $q=4,8,16$.
\item[\textsf{lz-index}\!\!]\footnote{\texttt{https://github.com/migumar2/uiHRDC/tree/master/uiHRDC/self-indexes/LZ}} is the only implementation of a Lempel-Ziv based
index \cite{KN12} that guarantees $O(z\log n)$ bits of space on a Lempel-Ziv
parse of $z$ phrases. We also use its
optimized implementation \cite{CFMN16}, which was shown to outperform 
\textsf{slp-index} both in space and time.
\item[\textsf{r-index}\!\!]\footnote{\texttt{https://github.com/nicolaprezza/r-index}} is the only implementation of a classical self-index (i.e., suffix-array based) using $O(r\log n)$ bits, where $r$
is the number of runs in the Burrows-Wheeler Transform of the text \cite{GNP18}.
\end{description}

We use six real repetitive collections from a repetitive corpus\footnote{\texttt{http://pizzachili.dcc.uchile.cl/repcorpus/real}}. 
Three of these collections contain DNA sequences extracted from differents sources: \textit{para} and \textit{cere} are extracted from the Saccharomyces Genome Resequencing Project\footnote{\texttt{http://www.sanger.ac.uk/Teams/Team71/durbin/sgrp}}, whereas \textit{influenza} is formed by DNA sequences of H.~Influenzae taken from the National Center for Biotechnology Information (NCBI)\footnote{\texttt{http://www.ncbi.nlm.nih.gov}}. 
Collection \textit{einstein.en} is formed by all version of the articles in English of Albert Einstein taken from Wikipedia. Collections \textit{kernel} and \textit{coreutils} are formed by all versions 5.x of the Coreutils package\footnote{\texttt{https://ftp.gnu.org/gnu/coreutils}} and all 1.0.x and 1.1.x versions of the Linux Kernel\footnote{\texttt{https://mirrors.edge.kernel.org/pub/linux/kernel}}, respectively.
Table~\ref{tab:ch_collections} lists their main characteristics.

\begin{table}[t]
    \centering
    \begin{tabular}{l|r|r|r|r|r|r}
          Collection                  & $n$~~~~~~~~ & $\sigma$~~\;\!   & $z$~~~~~~\,        & $r$~~~~~~~          & $G$ -- repair & $G$ -- proc~~ \\\hline
         \textit{para}              & ~429{,}265{,}758~  &  ~5~  & ~2{,}332{,}908~  & ~15{,}636{,}740~   & ~7{,}338{,}520~ 	& ~5{,}344{,}480~ \\ 
         \textit{cere}   	    & ~461{,}286{,}644~  &  ~5~  & ~1{,}700{,}859~  & ~11{,}574{,}641~   & ~5{,}780{,}080~  	& ~4{,}069{,}450~ \\
         \textit{influenza}         & ~154{,}808{,}555~  &  ~15~ & ~770{,}253~      & ~3{,}022{,}822~    & ~2{,}174{,}650~ 	& ~1{,}957{,}370~ \\ 
         \textit{einstein.en}~      & ~467{,}626{,}544~  & ~139~ & ~91{,}036~       & ~290{,}239~        & ~263{,}962~    	& ~212{,}903~  \\
         \textit{kernel}            & ~257{,}961{,}616~  & ~162~ & ~794{,}290~      & ~2{,}791{,}368~    & ~2{,}185{,}860~ 	& ~1{,}374{,}650~ \\
         \textit{coreutils}         & ~205{,}281{,}778~  & ~235~ & ~1{,}446{,}891~  & ~4{,}684{,}465~    & ~3{,}798{,}100~ 	& ~2{,}409{,}460~ \\
\hline
\end{tabular}
\ \\ \ \\
    \caption{Main characteristics of the collections: $n$ (size in bytes), 
$\sigma$ (alphabet size), $z$ (number of Lempel-Ziv phrases), $r$ (number of
runs in the BWT), and $G$ -- repair and $G$ -- proc (size of the RePair grammar
before and after applying the transformations of Section~\ref{sec:preproc},
respectively).}
    \label{tab:ch_collections}
\end{table}

Our times per occurrence are the average over 1000 patterns of length 10 extracted at random from each collection. Our extraction times average over 1000 
queries at random text positions.

\subsection{Locate Time}

\begin{figure*}[p]
	\includegraphics[width=0.49\textwidth]{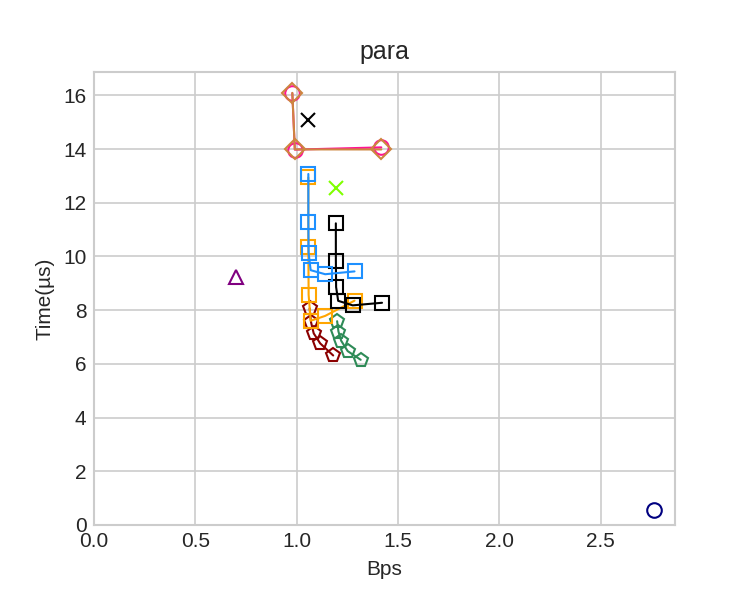}
	\includegraphics[width=0.49\textwidth]{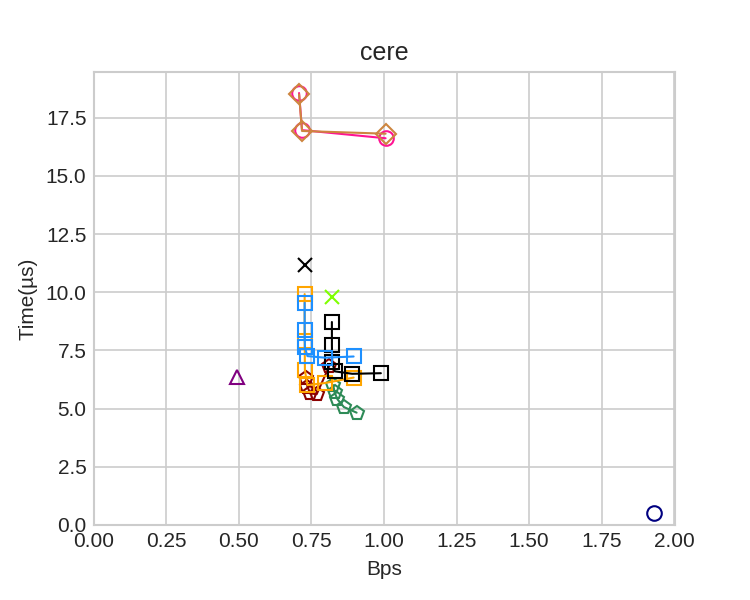}\\

\vspace*{-5mm}
	\includegraphics[width=0.49\textwidth]{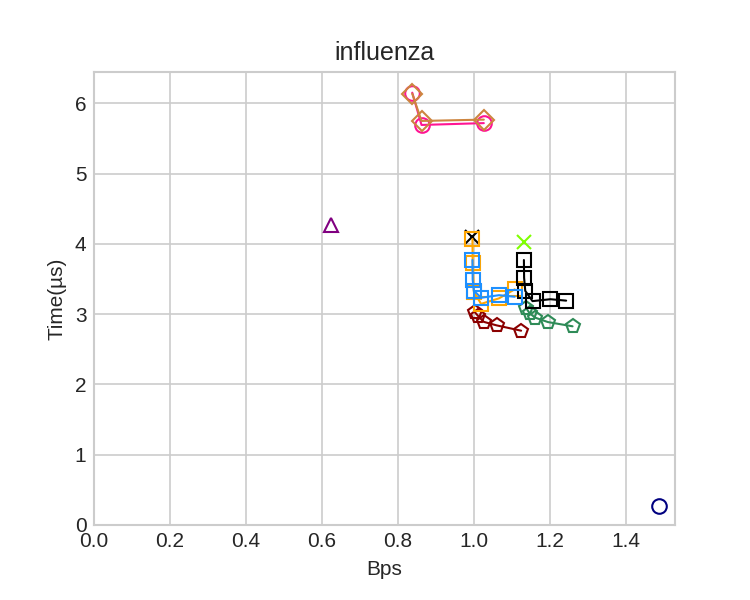}
	\includegraphics[width=0.49\textwidth]{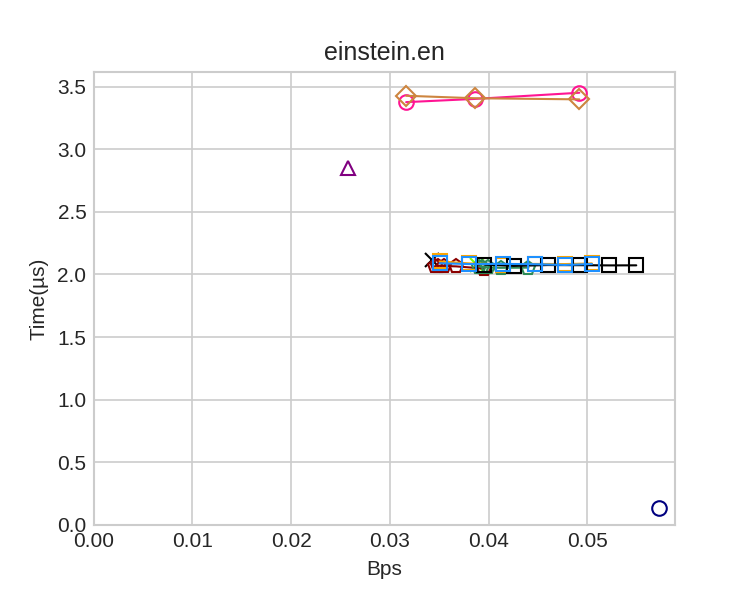}	\\

\vspace*{-5mm}
	\includegraphics[width=0.49\textwidth]{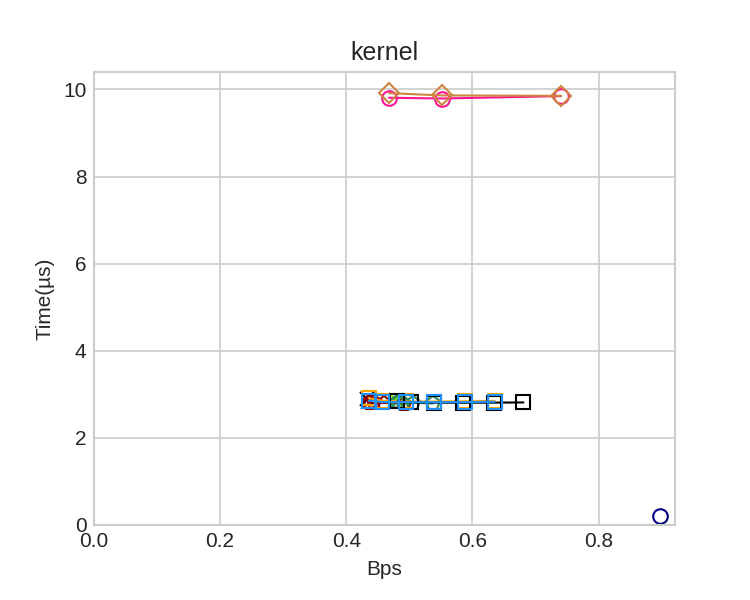}	
	\includegraphics[width=0.49\textwidth]{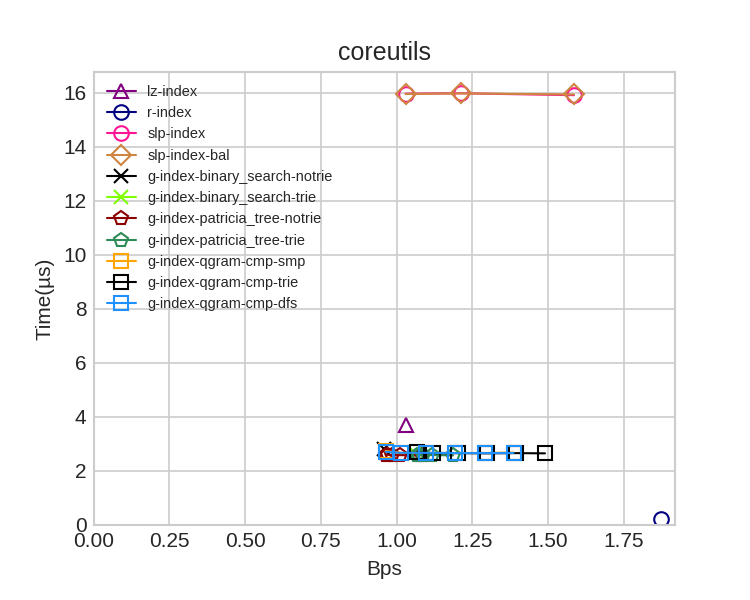}	\\

    \caption{Time-space tradeoffs for locating on different collections and indexes. The time is given in microseconds per occurrence and the space in bits per
symbol (bps).}
    \label{fig:locate}
\end{figure*}

Figure \ref{fig:locate} shows the space-time tradeoffs obtained for locating
patterns of length 10 over all the
indexes and parameter values on all the collections. We first discuss the 
results on our index and then compare it with the others.

In all cases, the use of Patricia trees with a sufficiently sparse sampling rate
can reach essentially the same space of the plain binary-search versions. Even
with the sparsest sampling rate (1 out of 64), the Patricia trees sharply
outperform the binary searches on the DNA alphabets, while making no difference
on the others. The rule samplings of the \textsf{qgram} versions also outperform 
binary searches on DNA alphabets without increasing the space, reaching the 
sweet point at value 8. Nevertheless, the Patricia trees make better use of the
space.

On the other hand, the use of the tries $T_S$ slightly increases the space 
while not providing a noticeable improvement in time (the exception is on the
binary-search versions of \textit{para}, but these lose anyway to the versions
using Patricia trees). As a result, we take 
\textsf{g-index-patricia\_tree-notrie} as the most convenient version of
our implementation, and we call it simply \textsf{g-index} henceforth.

The use of denser samplings yields an interesting space-time tradeoff for
\textsf{g-index} on DNA, which dominates a significant part of the Pareto curve.
On the other texts, it is better to use it with the sparsest sampling (or with
plain binary search), which dominates all the other alternatives on 
\textit{kernel} and \textit{coreutils}. With the sparsest sampling, 
\textsf{g-index} uses almost the same space of the previous grammar-based
index, \textsf{slp-index}, except on \textit{influenza}, where \textsf{g-index}
is 20\% larger. In exchange, \textsf{g-index} is up to 5 times faster than 
\textsf{slp-index}. There are almost no differences between \textsf{slp-index}
and \textsf{slp-index-bal}, which confirms that the grammar height does
not affect extraction time in practice.

Index \textsf{lz-index} outperforms \textsf{slp-index} in both space and time, 
as in previous work \cite{CFMN16}. While losing to \textsf{lz-index} in space 
is expected because $z \le G$ always holds, grammars allow for better methods 
to access the text. The index \textsf{slp-index} was, however, unable to take 
advantage of those methods to outperform \textsf{lz-index} in time. Now our
\textsf{g-index} does offer a space-time tradeoff, using more space than 
\textsf{lz-index}, but in exchange being faster; sometimes \textsf{lz-index}
is dominated in space and time. With sampling value 8, \textsf{g-index} is
50\%--65\% larger than \textsf{lz-index}, but 30\%--40\% faster on DNA
texts. With sampling value 64, \textsf{g-index} is from 10\% smaller to
25\% larger than \textsf{lz-index} and 20\%--40\% faster than it.

Finally, \textsf{r-index} is way faster than the others, but also way
larger (2--5 times larger than \textsf{lz-index}).

\begin{figure*}[t]
	\centering
    \includegraphics[width=0.49\textwidth]{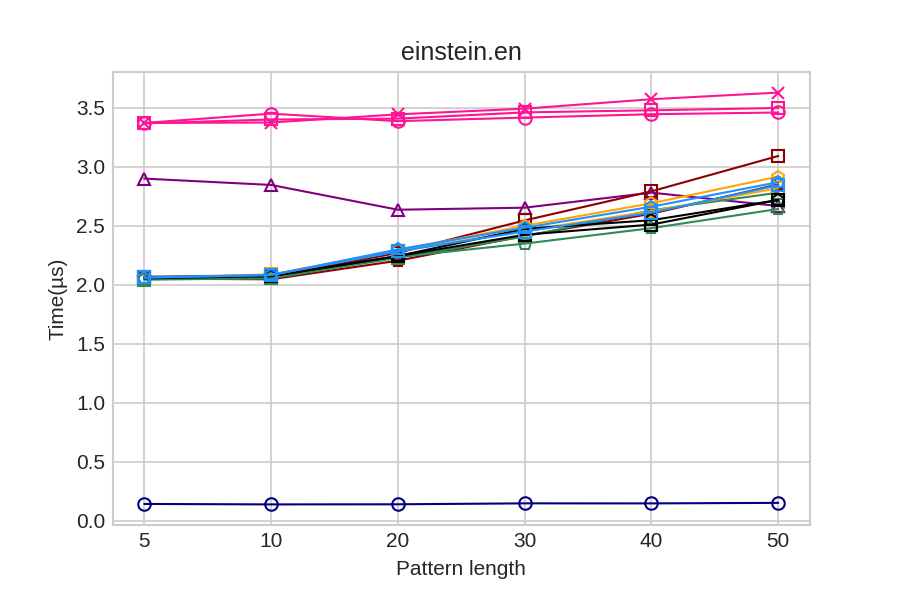} 
    \includegraphics[width=0.49\textwidth]{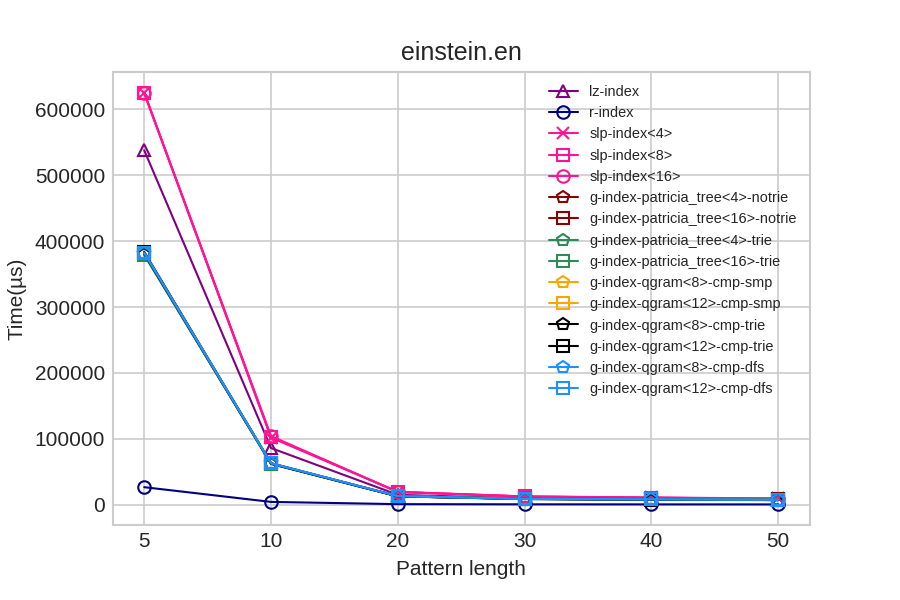} \\
    \caption{Locating time for increasing pattern lenghts on \textit{einstein.en}.
The time is given in microseconds per occurrence on the left, and per 
pattern symbol on the right. Only some representative indexes are shown on the
right plot.}
    \label{fig:locate-len}
\end{figure*}

Figure~\ref{fig:locate-len} shows how the locate time evolves with the pattern
length $m$ on \textit{einstein.en}. We still include the different variants of 
our index in this plot (the numbers in brackets are the sample values), and 
consider pattern lengths 5, 10, 20, 30, 40, and 50.

The left plot shows that, while \textsf{g-index} is way faster than 
\textsf{slp-index} and \textsf{lz-index} on this text for small $m$, all 
the \textsf{g-index} variants slow down as $m$ increases, eventually losing to 
\textsf{lz-index} for long enough patterns. 
The most important lengths, however, where a large number of occurrences are
found, are the short ones. The total query times are much less significant for
long patterns, as shown on the right plot.

\subsection{Extraction Time}

\begin{figure*}[p]
	\includegraphics[width=0.49\textwidth]{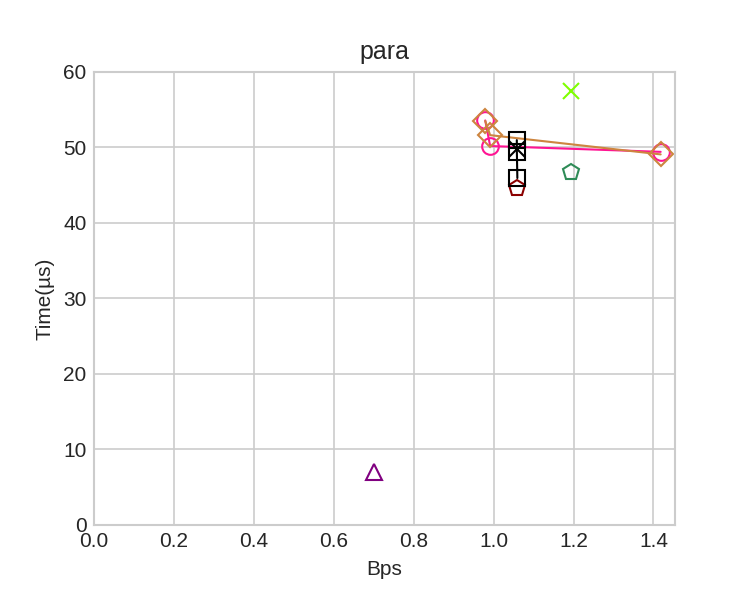}
	\includegraphics[width=0.49\textwidth]{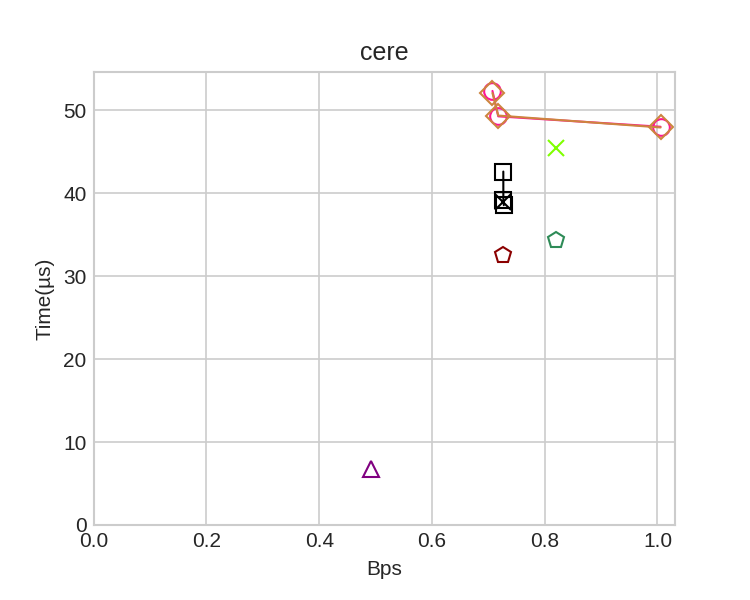}\\

\vspace*{-5mm}
	\includegraphics[width=0.49\textwidth]{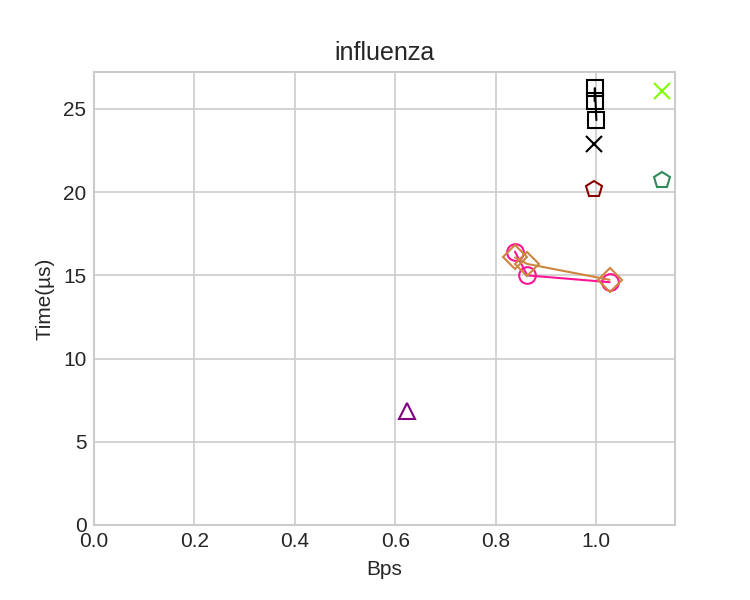}
	\includegraphics[width=0.49\textwidth]{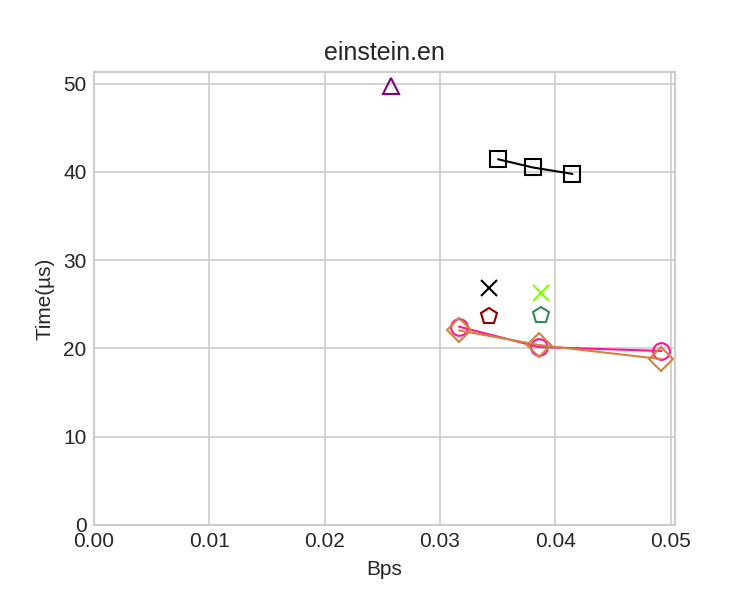}	\\

\vspace*{-5mm}
	\includegraphics[width=0.49\textwidth]{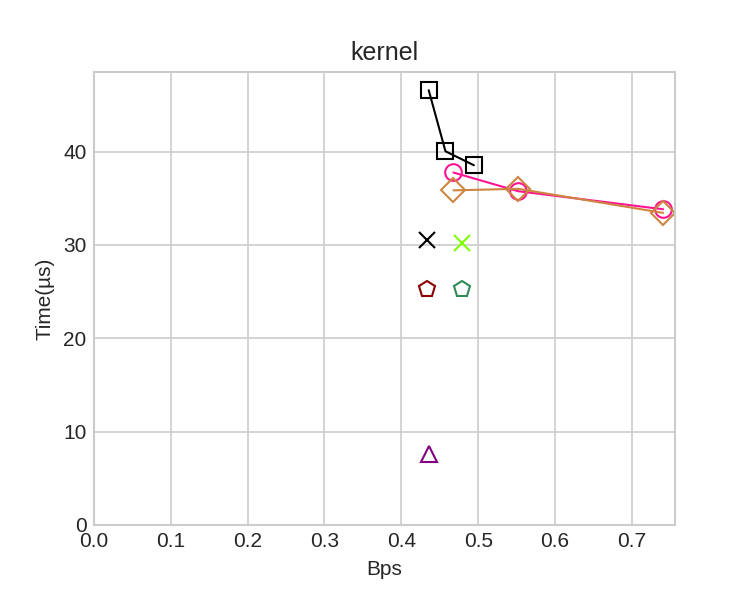}	
	\includegraphics[width=0.49\textwidth]{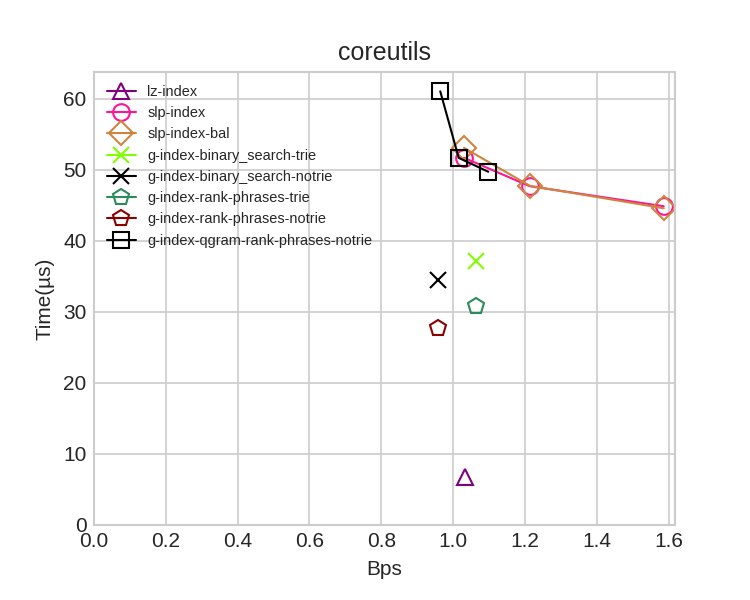}	\\

    \caption{Time-space tradeoffs for extracting on different collections and indexes. The time is given in microseconds per extracted symbol and the space in bits per
symbol (bps).}
    \label{fig:display}
\end{figure*}

Figure \ref{fig:display} shows the time per extracted symbol of the 
different indexes and collections, when extracting 10 consecutive 
text symbols. The \textsf{r-index} is excluded because it does not support 
this operation. Note that the extraction in our \textsf{g-index} is independent
of whether or not we use binary search or Patricia trees. The variant that
continues with \textsf{binary\_search} descends from the root symbol, binary
searching the children, to reach the desired substring to extract. Instead,
the variant that continues with \textsf{rank\_phrases} uses $\rank$ on the
bitmap $L$ (Section~\ref{sec:extractarbitrary}), so as to find faster the
phrases to be expanded. The suffixes \textsf{trie} and \textsf{notrie} refer
again to using or not the structure for extracting prefixes and suffixes in
linear time (for the phrases that are not completely contained in the area to
extract). Finally, if \textsf{qgram} follows \textsf{g-index}, we use
the $q$-grams to speed up extraction, with lengths $2,4,6,8$.

As it can be seen, the best \textsf{g-index} variant, both in space and time,
is usually \textsf{g-index-rank\_phrases-notrie}. It is also apparent that 
\textsf{lz-index} excells in extraction, being 3--5 times faster than every 
\textsf{g-index} variant (except on \textit{einstein.en}, the most repetitive 
collection, where the $\bar{h}$ value of the Lempel-Ziv parsing is very high). 
On the other hand, our best \textsf{g-index} variant outperforms 
\textsf{slp-index} in time by 20\%--100\% within similar space. 
The exceptions are \textit{influenza} and \textit{einstein.en}, where 
\textsf{slp-index} is 10\%--30\% faster. 

\section{Conclusions }

We have presented the first compressed text index based on arbitrary
context-free grammars whose time per retrieved occurrence is logarithmic,
independently of the grammar height. Given a text $T[1..n]$ represented by a
grammar of size $G$, our index uses $O(G\log n)$ bits of space and returns the 
$occ$ occurrences of a pattern $P[1..m]$ in time $O((m^2 + occ)\log G)$. We 
implemented our index and compared it with various alternatives in the 
literature, showing that it is practical and offers relevant space/time 
tradeoffs.

The most interesting open theoretical question is whether it is possible to
obtain $O(G\log g)$ bits, as grammar-based compressors could reach, instead of
$O(G\log n)$, since in some text families we have $g\le G=O(\log n)$. This 
term owes
to storing the lengths of the expansions of the nonterminals in bitmap $L$. 
We tried storing these lengths in the nonterminals, instead, and sampling the
nonterminals that would store lengths. Finding a suitable sampling on the
grammar DAG, however, is related to finding minimum cuts in graphs \cite{AHK04},
which is not easy.

With respect to practical results, an interesting research direction would be
to obtain practical implementations of recent proposals that reduce the 
$O(m^2)$ term in the search complexity \cite{GGKNP14,BEGV18,CE18}. Some of 
those methods, however, have a penalty factor of $O(\log\log z)$ or
$O(\log(n/z))$ in the space, which is far from negligible, and thus they are
unlikely to be competitive in space. The index of Christiansen and Ettienne
\cite{CE18}, on the other hand, builds on a grammar and looks more promising.
They manage to ensure that $P$ needs be cut into only $O(\log m)$ places to
spot all the primary occurrences, which reduces the $O(m^2)$ term to 
$O(m\log m)$. For this to hold, however, the grammar must be of a special 
type called locally-consistent.
In our experience, RePair outperforms in space, by a wide margin, all the
other grammar construction algorithms, including those that offer guarantees
of the form $G = O(z\log(n/z))$. It is therefore unclear which is the price
to pay in space in order to use a specific type of grammar.

\bibliographystyle{alpha}
\bibliography{paper}

\newcommand{\etalchar}[1]{$^{#1}$}
\begin{thebibliography}{CFMPN16}

\bibitem[AB92]{AB92}
A.~Amir and G.~Benson.
\newblock Efficient two-dimensional compressed matching.
\newblock In {\em Proc. 2nd Data Compression Conference (DCC)}, pages 279--288,
  1992.

\bibitem[AHK04]{AHK04}
S.~Arora, E.~Hazan, and S.~Kale.
\newblock {$O(\sqrt{\log n})$} approximation to {SPARSEST CUT} {$O(n^2)$} in
  time.
\newblock In {\em Proc. 45th Annual Symposium on Foundations of Computer
  Science (FOCS)}, pages 238--247, 2004.

\bibitem[ANS12]{ANS10}
D.~Arroyuelo, G.~Navarro, and K.~Sadakane.
\newblock Stronger {L}empel-{Z}iv based compressed text indexing.
\newblock {\em Algorithmica}, 62(1):54--101, 2012.

\bibitem[BCG{\etalchar{+}}15]{BCGPR15}
D.~Belazzougui, F.~Cunial, T.~Gagie, N.~Prezza, and M.~Raffinot.
\newblock Composite repetition-aware data structures.
\newblock In {\em Proc. 26th Annual Symposium on Combinatorial Pattern Matching
  (CPM)}, LNCS 9133, pages 26--39, 2015.

\bibitem[BCN13]{BCN13}
J.~Barbay, F.~Claude, and G.~Navarro.
\newblock Compact binary relation representations with rich functionality.
\newblock {\em Information and Computation}, 232:19--37, 2013.

\bibitem[BDM{\etalchar{+}}05]{BDM+05}
D.~Benoit, E.~Demaine, J.~I. Munro, R.~Raman, V.~Raman, and S.~Srinivasa Rao.
\newblock Representing trees of higher degree.
\newblock {\em Algorithmica}, 43(4):275--292, 2005.

\bibitem[BEGV18]{BEGV18}
P.~Bille, M.~B. Ettienne, I.~L. G{\o}rtz, and H.~W. Vildh{\o}j.
\newblock Time-space trade-offs for {Lempel-Ziv} compressed indexing.
\newblock {\em Theoretical Computer Science}, 713:66--77, 2018.

\bibitem[BGG{\etalchar{+}}14]{BGGMS14}
D.~Belazzougui, T.~Gagie, S.~Gog, G.~Manzini, and J.~Sir{\'e}n.
\newblock Relative {FM}-indexes.
\newblock In {\em Proc. 21st International Symposium on String Processing and
  Information Retrieval (SPIRE)}, LNCS 8799, pages 52--64, 2014.

\bibitem[BLR{\etalchar{+}}15]{BLRSRW15}
P.~Bille, G.~M. Landau, R.~Raman, K.~Sadakane, S.~S. Rao, and O.~Weimann.
\newblock Random access to grammar-compressed strings and trees.
\newblock {\em SIAM Journal on Computing}, 44(3):513--539, 2015.

\bibitem[BN15]{BN14}
D.~Belazzougui and G.~Navarro.
\newblock Optimal lower and upper bounds for representing sequences.
\newblock {\em ACM Transactions on Algorithms}, 11(4):article 31, 2015.

\bibitem[BPT15]{BPT15}
D.~Belazzougui, S.~J. Puglisi, and Y.~Tabei.
\newblock Access, rank, select in grammar-compressed strings.
\newblock In {\em Proc. 23rd Annual European Symposium on Algorithms (ESA)},
  LNCS 9294, pages 142--154, 2015.

\bibitem[CE18]{CE18}
A.~R. Christiansen and M.~B. Ettienne.
\newblock Compressed indexing with signature grammars.
\newblock In {\em Proc. 13th Latin American Symposium on Theoretical
  Informatics (LATIN)}, pages 331--345, 2018.

\bibitem[CFMPN10]{CFMPN10}
F.~Claude, A.~Fari{\~n}a, M.~Mart{\'{\i}}nez-Prieto, and G.~Navarro.
\newblock Compressed $q$-gram indexing for highly repetitive biological
  sequences.
\newblock In {\em Proc. 10th IEEE Conference on Bioinformatics and
  Bioengineering (BIBE)}, 2010.

\bibitem[CFMPN16]{CFMN16}
F.~Claude, A.~Fari{\~n}a, M.~Mart{\'{\i}}nez-Prieto, and G.~Navarro.
\newblock Universal indexes for highly repetitive document collections.
\newblock {\em Information Systems}, 61:1--23, 2016.

\bibitem[Cla96]{Cla96}
D.~Clark.
\newblock {\em Compact Pat Trees}.
\newblock PhD thesis, University of Waterloo, 1996.

\bibitem[CLL{\etalchar{+}}05]{CLLPPSS05}
M.~Charikar, E.~Lehman, D.~Liu, R.~Panigrahy, M.~Prabhakaran, A.~Sahai, and
  A.~Shelat.
\newblock The smallest grammar problem.
\newblock {\em IEEE Transactions on Information Theory}, 51(7):2554--2576,
  2005.

\bibitem[CLP11]{CLP11}
T.~M. Chan, K.~G. Larsen, and M.~P{\u{a}}tra\c{s}cu.
\newblock Orthogonal range searching on the {RAM}, revisited.
\newblock In {\em Proc. 27th ACM Symposium on Computational Geometry (SoCG)},
  pages 1--10, 2011.

\bibitem[CN10]{CN11}
F.~Claude and G.~Navarro.
\newblock Self-indexed grammar-based compression.
\newblock {\em Fundamenta Informaticae}, 111(3):313--337, 2010.

\bibitem[CN12]{CN12}
F.~Claude and G.~Navarro.
\newblock Improved grammar-based compressed indexes.
\newblock In {\em Proc. 19th International Symposium on String Processing and
  Information Retrieval (SPIRE)}, LNCS 7608, pages 180--192, 2012.

\bibitem[CRA76]{CRA76}
C.~Cook, A.~Rosenfeld, and A.~Aronson.
\newblock Grammatical inference by hill climbing.
\newblock {\em Information Science}, 10:59--–80, 1976.

\bibitem[DJSS14]{DJSS14}
H.~H. Do, J.~Jansson, K.~Sadakane, and W.-K. Sung.
\newblock Fast relative {Lempel-Ziv} self-index for similar sequences.
\newblock {\em Theoretical Computer Science}, 532:14--30, 2014.

\bibitem[Eli74]{Eli74}
P.~Elias.
\newblock Efficient storage and retrieval by content and address of static
  files.
\newblock {\em Journal of the ACM}, 21:246--260, 1974.

\bibitem[Fan71]{Fan71}
R.~Fano.
\newblock On the number of bits required to implement an associative memory.
\newblock Memo 61, Computer Structures Group, Project MAC, Massachusetts, 1971.

\bibitem[FM05]{FM05}
P.~Ferragina and G.~Manzini.
\newblock Indexing compressed texts.
\newblock {\em Journal of the ACM}, 52(4):552--581, 2005.

\bibitem[GGK{\etalchar{+}}12]{GGKNP12}
T.~Gagie, P.~Gawrychowski, J.~K{\"{a}}rkk{\"{a}}inen, Y.~Nekrich, and S.~J.
  Puglisi.
\newblock A faster grammar-based self-index.
\newblock In {\em Proc. 6th International Conference on Language and Automata
  Theory and Applications (LATA)}, LNCS 7183, pages 240--251, 2012.

\bibitem[GGK{\etalchar{+}}14]{GGKNP14}
T.~Gagie, P.~Gawrychowski, J.~K{\"{a}}rkk{\"{a}}inen, Y.~Nekrich, and S.~J.
  Puglisi.
\newblock {LZ77}-based self-indexing with faster pattern matching.
\newblock In {\em Proc. 11th Latin American Theoretical Informatics Symposium
  (LATIN)}, LNCS 8392, pages 731--742, 2014.

\bibitem[GKPS05]{GKPS05}
L.~Gasieniec, R.~Kolpakov, I.~Potapov, and P.~Sant.
\newblock Real-time traversal in grammar-based compressed files.
\newblock In {\em Proc. 15th Data Compression Conference (DCC)}, page 458,
  2005.

\bibitem[GMR06]{GMR06}
A.~Golynski, J.~I. Munro, and S.~Rao.
\newblock Rank/select operations on large alphabets: a tool for text indexing.
\newblock In {\em Proc. 17th Annual ACM-SIAM Symposium on Discrete Algorithms
  (SODA)}, pages 368--373, 2006.

\bibitem[GNP18]{GNP18}
T.~Gagie, G.~Navarro, and N.~Prezza.
\newblock Optimal-time text indexing in {BWT}-runs bounded space.
\newblock In {\em Proc. 29th Annual ACM-SIAM Symposium on Discrete Algorithms
  (SODA)}, pages 1459--1477, 2018.

\bibitem[HLR16]{HLR16}
D.~Hucke, M.~Lohrey, and C.~P. Reh.
\newblock The smallest grammar problem revisited.
\newblock In {\em Proc. 23rd International Symposium on String Processing and
  Information Retrieval}, LNCS 9954, pages 35--49, 2016.

\bibitem[Jez15]{Jez15}
A.~Jez.
\newblock Approximation of grammar-based compression via recompression.
\newblock {\em Theoretical Computer Science}, 592:115--134, 2015.

\bibitem[Jez16]{Jez16}
A.~Jez.
\newblock A really simple approximation of smallest grammar.
\newblock {\em Theoretical Computer Science}, 616:141--150, 2016.

\bibitem[K{\"a}r99]{Kar99}
J.~K{\"a}rkk{\"a}inen.
\newblock {\em Repetition-Based Text Indexing}.
\newblock PhD thesis, U. Helsinki, Finland, 1999.

\bibitem[KMS{\etalchar{+}}03]{KMSTSA03}
T.~Kida, T.~Matsumoto, Y.~Shibata, M.~Takeda, A.~Shinohara, and S.~Arikawa.
\newblock Collage system: a unifying framework for compressed pattern matching.
\newblock {\em Theoretical Computer Science}, 298(1):253--272, 2003.

\bibitem[KN13]{KN12}
S.~Kreft and G.~Navarro.
\newblock On compressing and indexing repetitive sequences.
\newblock {\em Theoretical Computer Science}, 483:115--133, 2013.

\bibitem[KY00]{KY00}
J.~Kieffer and E.-H. Yang.
\newblock Grammar-based codes: A new class of universal lossless source codes.
\newblock {\em IEEE Transactions on Information Theory}, 46(3):737--754, 2000.

\bibitem[LM00]{LM00}
J.~Larsson and A.~Moffat.
\newblock Off-line dictionary-based compression.
\newblock {\em Proc. IEEE}, 88(11):1722--1732, 2000.

\bibitem[LZ76]{LZ76}
A.~Lempel and J.~Ziv.
\newblock On the complexity of finite sequences.
\newblock {\em IEEE Transactions on Information Theory}, 22(1):75--81, 1976.

\bibitem[MNSV10]{MNSV09}
V.~M{\"a}kinen, G.~Navarro, J.~Sir{\'e}n, and N.~V{\"a}lim{\"a}ki.
\newblock Storage and retrieval of highly repetitive sequence collections.
\newblock {\em Journal of Computational Biology}, 17(3):281--308, 2010.

\bibitem[Mor68]{Mor68}
D.~Morrison.
\newblock {PATRICIA} -- practical algorithm to retrieve information coded in
  alphanumeric.
\newblock {\em Journal of the ACM}, 15(4):514--534, 1968.

\bibitem[MRRR12]{MRRR12}
J.~I. Munro, R.~Raman, V.~Raman, and S.~S. Rao.
\newblock Succinct representations of permutations and functions.
\newblock {\em Theoretical Computer Science}, 438:74--88, 2012.

\bibitem[Nav12]{Nav12}
G.~Navarro.
\newblock Indexing highly repetitive collections.
\newblock In {\em Proc. 23rd International Workshop on Combinatorial Algorithms
  (IWOCA)}, LNCS 7643, pages 274--279, 2012.

\bibitem[NM07]{NM07}
G.~Navarro and V.~M{\"a}kinen.
\newblock Compressed full-text indexes.
\newblock {\em ACM Computing Surveys}, 39(1):article 2, 2007.

\bibitem[NMWM94]{NMWM04}
C.~Nevill-Manning, I.~Witten, and D.~Maulsby.
\newblock Compression by induction of hierarchical grammars.
\newblock In {\em Proc. 4th Data Compression Conference (DCC)}, pages 244--253,
  1994.

\bibitem[NP19]{NP18}
G.~Navarro and N.~Prezza.
\newblock Universal compressed text indexing.
\newblock {\em Theoretical Computer Science}, 762:41--50, 2019.

\bibitem[NPC{\etalchar{+}}13]{NPCHIMP13}
J.~C. Na, H.~Park, M.~Crochemore, J.~Holub, C.~S. Iliopoulos, L.~Mouchard, and
  K.~Park.
\newblock Suffix tree of alignment: An efficient index for similar data.
\newblock In {\em Proc. 24th International Workshop on Combinatorial Algorithms
  (IWOCA)}, LNCS 8288, pages 337--348, 2013.

\bibitem[NPL{\etalchar{+}}13]{NPLHLMP13}
J.~C. Na, H.~Park, S.~Lee, M.~Hong, T.~Lecroq, L.~Mouchard, and K.~Park.
\newblock Suffix array of alignment: {A} practical index for similar data.
\newblock In {\em Proc. 20th International Symposium on String Processing and
  Information Retrieval (SPIRE)}, LNCS 8214, pages 243--254, 2013.

\bibitem[NS14]{NS14}
G.~Navarro and K.~Sadakane.
\newblock Fully-functional static and dynamic succinct trees.
\newblock {\em ACM Transactions on Algorithms}, 10(3):article 16, 2014.

\bibitem[RO08]{RO08}
L.~Russo and A.~Oliveira.
\newblock A compressed self-index using a {Z}iv-{L}empel dictionary.
\newblock {\em Information Retrieval}, 11(4):359--388, 2008.

\bibitem[RRR07]{RRR07}
R.~Raman, V.~Raman, and S.~S. Rao.
\newblock Succinct indexable dictionaries with applications to encoding {\it
  k}-ary trees, prefix sums and multisets.
\newblock {\em ACM Transactions on Algorithms}, 3(4):article 43, 2007.

\bibitem[Ryt03]{Ryt03}
W.~Rytter.
\newblock Application of {L}empel-{Z}iv factorization to the approximation of
  grammar-based compression.
\newblock {\em Theoretical Computer Science}, 302(1-3):211--222, 2003.

\bibitem[SS82]{SS82}
J.~A. Storer and T.~G. Szymanski.
\newblock Data compression via textual substitution.
\newblock {\em Journal of the {ACM}}, 29(4):928--951, 1982.

\bibitem[Sto77]{Sto77}
J.~A. Storer.
\newblock {NP}-completeness results concerning data compression.
\newblock Technical Report 234, Department of Electrical Engineering and
  Computer Science, Princeton University, 1977.

\bibitem[VY13]{VY13}
E.~Verbin and W.~Yu.
\newblock Data structure lower bounds on random access to grammar-compressed
  strings.
\newblock In {\em Proc. 24th Annual Symposium on Combinatorial Pattern Matching
  (CPM)}, LNCS 7922, pages 247--258, 2013.

\bibitem[ZL78]{ZL78}
J.~Ziv and A.~Lempel.
\newblock Compression of individual sequences via variable length coding.
\newblock {\em IEEE Transactions on Information Theory}, 24(5):530--536, 1978.

\end{thebibliography}


\end{document}